\documentclass[twocolumn]{aastex63}

\usepackage{array}
\usepackage{amsmath}
\usepackage{xcolor}
\usepackage{multirow}
\usepackage{booktabs}
\usepackage{soul}

\def\jsunit{$\,$km$\,$s$^{-1}\,$pc}
\def\Jsunit{$\,M_{\odot}\,$km$\,$s$^{-1}\,$pc}
\def\au{$\,$au}
\def\mdotunit{$\,M_{\odot}\,$yr$^{-1}$}
\def\msun{$\,M_{\odot}$}

\def\NNHp{N$_{2}$H$^{+}$}
\def\COa{C$^{18}$O} 
\def\COan{C$^{18}$O$\,(J=2\rightarrow1)$} 
\def\COb{$^{13}$CO} 
\def\CObn{$^{13}$CO$\,(J=2\rightarrow1)$} 
\def\COc{$^{12}$CO} 
\def\COcn{$^{12}$CO$\,(J=2\rightarrow1)$} 
\def\SO{SO}
\def\SOn{SO$\,(J=5_{6}\rightarrow4_{5})$}
\def\NNDp{N$_{2}$D$^{+}$}

\def\continuum{1.3$\,$mm continuum}
\def\HCOp{HCO$^{+}$}
\def\HCOpn{HCO$^{+}\,(J=3\rightarrow2)$}

\hypersetup{linkcolor=blue,citecolor=blue,filecolor=blue,urlcolor=blue}

\submitjournal{ApJ}

\shorttitle{Accretion Flows or Outflow Cavities?}
\shortauthors{Thieme et al.}

\begin{document}

\title{Accretion Flows or Outflow Cavities? 
Uncovering the Gas Dynamics around Lupus 3-MMS}


\author[0000-0003-0334-1583]{Travis J. Thieme}
\affiliation{Institute of Astronomy, National Tsing Hua University, No. 101, Section 2, Kuang-Fu Road, Hsinchu 30013, Taiwan}
\affiliation{Center for Informatics and Computation in Astronomy, National Tsing Hua University, No. 101, Section 2, Kuang-Fu Road, Hsinchu 30013, Taiwan}
\affiliation{Department of Physics, National Tsing Hua University, No. 101, Section 2, Kuang-Fu Road, Hsinchu 30013, Taiwan}

\author[0000-0001-5522-486X]{Shih-Ping Lai}
\affiliation{Institute of Astronomy, National Tsing Hua University, No. 101, Section 2, Kuang-Fu Road, Hsinchu 30013, Taiwan}
\affiliation{Center for Informatics and Computation in Astronomy, National Tsing Hua University, No. 101, Section 2, Kuang-Fu Road, Hsinchu 30013, Taiwan}
\affiliation{Department of Physics, National Tsing Hua University, No. 101, Section 2, Kuang-Fu Road, Hsinchu 30013, Taiwan}
\affiliation{Academia Sinica Institute of Astronomy and Astrophysics, P.O. Box 23-141, Taipei 10617, Taiwan}

\author[0000-0002-6868-4483]{Sheng-Jun Lin}
\affiliation{Institute of Astronomy, National Tsing Hua University, No. 101, Section 2, Kuang-Fu Road, Hsinchu 30013, Taiwan}
\affiliation{Center for Informatics and Computation in Astronomy, National Tsing Hua University, No. 101, Section 2, Kuang-Fu Road, Hsinchu 30013, Taiwan}
\affiliation{Department of Physics, National Tsing Hua University, No. 101, Section 2, Kuang-Fu Road, Hsinchu 30013, Taiwan}

\author{Pou-Ieng Cheong}
\affiliation{Institute of Astronomy, National Tsing Hua University, No. 101, Section 2, Kuang-Fu Road, Hsinchu 30013, Taiwan}
\affiliation{Department of Physics, National Tsing Hua University, No. 101, Section 2, Kuang-Fu Road, Hsinchu 30013, Taiwan}

\author[0000-0002-3024-5864]{Chin-Fei Lee}
\affiliation{Academia Sinica Institute of Astronomy and Astrophysics, P.O. Box 23-141, Taipei 10617, Taiwan}
\affiliation{Graduate Institute of Astronomy and Astrophysics, National Taiwan University, No. 1, Sec. 4, Roosevelt Road, Taipei 10617, Taiwan }

\author[0000-0003-1412-893X]{Hsi-Wei Yen}
\affiliation{Academia Sinica Institute of Astronomy and Astrophysics, P.O. Box 23-141, Taipei 10617, Taiwan}

\author[0000-0002-7402-6487]{Zhi-Yun Li}
\affiliation{Department of Astronomy, University of Virginia, 530 McCormick Road, Charlottesville, Virginia 22904-4325, USA}

\author[0000-0003-3581-1834]{Ka Ho Lam}
\affiliation{Department of Astronomy, University of Virginia, 530 McCormick Road, Charlottesville, Virginia 22904-4325, USA}

\author[0000-0002-5359-8072]{Bo Zhao}
\affiliation{Department of Physics and Astronomy, McMaster University, 1280 Main Street, West Hamilton, Ontario L8S 4L8, Canada}

\begin{abstract}
Understanding how material accretes onto the rotationally supported disk from the surrounding envelope of gas and dust in the youngest protostellar systems is important for describing how disks are formed. Magnetohydrodynamic simulations of magnetized, turbulent disk formation usually show spiral-like streams of material (accretion flows) connecting the envelope to the disk. 
However, accretion flows in these early stages of protostellar formation still remain poorly characterized due to their low intensity and possibly some extended structures are disregarded as being part of the outflow cavity. 
We use ALMA archival data of a young Class 0 protostar, Lupus 3-MMS, to uncover four extended accretion flow-like structures in \COa\ that follow the edges of the outflows.
We make various types of position-velocity cuts to compare with the outflows and find the extended structures are not consistent with the outflow emission, but rather more consistent with a simple infall model. 
We then use a dendrogram algorithm to isolate five sub-structures in position-position-velocity space.
Four out of the five sub-structures fit well ($>$95\%) with our simple infall model, with specific angular momenta between $2.7-6.9\times10^{-4}$\jsunit\ and mass-infall rates of $0.5-1.1\times10^{-6}$\mdotunit.
Better characterization of the physical structure in the supposed ``outflow-cavities'' is important to disentangle the true outflow cavities and accretion flows. 
\end{abstract}

\keywords{Accretion (14) --- Circumstellar disks (235) --- Observational astronomy (1145) --- Protostars (1302) --- Radio astronomy (1338) --- Radio interferometry (1346) --- Star formation (1569) --- Stellar accretion (1578) --- Stellar accretion disks (1579) --- Young stellar objects (1834)}

\section{Introduction} \label{sec:introduction}
Rotationally supported disks around the youngest protostars are essential for transporting the angular momentum of infalling material from the envelope to the protostar and for the development of proto-planetary systems. 
Observations of molecular lines trace the Keplerian rotation of these disks around Class 0 protostars and confirm that disks form in the earliest stages of star formation when the protostar is still deeply embedded in its parent core \citep[e.g.,][]{Tobin2012Natur.492...83T,Murillo2013A&A...560A.103M,Ohashi2014ApJ...796..131O,Lee2014ApJ...786..114L,Codella2014A&A...568L...5C,Aso2017ApJ...849...56A,Yen2017ApJ...834..178Y}. 
The size of these gaseous disks in the Class 0 stage range from about 20\au\ to 150\au\ around protostellar masses less than 0.5\msun.
As the protostar evolves into the later Class I phase, the observed disk radii ranges between 50\au\ to $700$\au\ around protostellar masses less than 2.5\msun \citep[e.g.,][]{Lommen2008A&A...481..141L,Harsano2014A&A...562A..77H,Yen2015ApJ...799..193Y}. 
The wide range of disk radii show that their formation is complicated and can have a variety of outcomes.
In order to better understand the formation and growth of these young disks, it is important to study how material accretes onto the disk from the surrounding envelope. 

A rotationally supported disk (hereafter, RSD) is expected to form around the protostar due to the conservation of angular momentum \citep[e.g.,][]{Ulrich1976ApJ...210..377U}. 
In this simplified model of axisymmetric gravitational collapse, the infalling material follow free-fall parabolic trajectories that collide with the disk under the assumption that the central object can be approximated as a point source \citep[CMU Model:][]{Ulrich1976ApJ...210..377U,Cassen1981Icar...48..353C}. 
Under the `inside-out' assumption \citep{Shu1977ApJ...214..488S}, the disk radius is then expected to grow as $\propto M_{\mathrm{sd}}^{3}$, where $M_{\mathrm{sd}}$ is the total mass of the star+disk system \citep{Terebey1984ApJ...286..529T}.
While this CMU model can describe the kinematics of the infalling material, it is possible for the material to fall in from every direction around the disk in a globalized collapse scenario, or in filamentary-like streams due to asymmetric density structures in the envelope \citep{Tobin2012ApJ...748...16T}. 
On the other hand, magnetohydrodynamic (MHD) simulations of dense cores show spiral-like streams of material connecting the envelope to the disk in magnetized turbulent cores \citep{Li2014ApJ...793..130L,Seifried2015MNRAS.446.2776S}.
These models show that the accretion flow structures, as well as the formation of a rotationally supported disk, can vary depending on the levels of turbulence and magnetic fields in the system. 
In the ideal MHD limit, if the magnetic field is strongly coupled with the rotating gas within the protostellar core, it becomes twisted and can effectively carry away angular momentum from the central region and inhibit the formation of protostellar disks \citep[\textit{Magnetic Braking Catastrophe:} ][]{Mellon2008ApJ...681.1356M,Hennebelle2008A&A...477....9H}.
Several mechanisms have been proposed to weaken the magnetic braking effect and allow for the formation of protostellar disks of comparable sizes to those previously observed, including turbulence \citep[e.g.,][]{Joos2013A&A...554A..17J,Seifried2013MNRAS.432.3320S}, misalignment between the magnetic field and rotation axis \citep[e.g.,][]{Hennebelle2009A&A...506L..29H, Li2013ApJ...774...82L}, non-ideal MHD effects \citep[e.g.,][]{Li2011ApJ...738..180L,  Wurster2016MNRAS.457.1037W}, micro-physics \citep[e.g.,][]{Zhao2016MNRAS.460.2050Z,Wurster2018MNRAS.476.2063W,Kuffmeier2020A&A...639A..86K}, and various combinations of these scenarios \citep[e.g.,][]{Tsukamoto2018ApJ...868...22T, Lam2019MNRAS.489.5326L, Wurster2020MNRAS.495.3795W}.
Simulations with these properties to overcome magnetic braking almost always show accretion flows at various stages of their time evolution.
Thus, looking for these structures around young protostars and comparing with models can help us to better understand the disk formation process.

Observationally, these accretion flow-like structures have been reported in a few Class I protostars (L1489-IRS: \citealt[][]{Yen2014ApJ...793....1Y}; HL Tau: \citealt[][]{Yen2017AA...608A.134Y,Yen2019ApJ...880...69Y}) and Class 0 protostars (VLA1623: \citealt{CheongThesis2018}, \citealt{Hsieh2020ApJ...894...23H}; IRAS 03292+3039: \citealt[][]{Pineda2020NatAs.tmp..151P}).
In L1489-IRS, \citet{Yen2014ApJ...793....1Y} found two infalling flows with lengths of 5000\au\ (redshifted flow) and 2000\au\ (blueshifted flow) in \COan.
They find these flows are consistent with the CMU model using a centrifugal radius of 300\au.
They estimate the mass infalling rate to be $4-7\times10^{-7}$\mdotunit.
In HL Tau, \citet[][]{Yen2017AA...608A.134Y,Yen2019ApJ...880...69Y} found two infalling flows in \CObn, one of which is also seen in \HCOpn. 
The \COb\ accretion flows can be matched with the CMU model with a centrifugal radius of 100\au\ and a specific angular momentum of $1.9 \times 10^{-3}$\jsunit. 
They estimate the mass infalling rate to be $2.2\times10^{-6}$\mdotunit. 
In VLA1623, \citet{CheongThesis2018} identify two accretion flows in \COan\ using a clump finding ``dendrogram" algorithm \citep{Rosolowsky2008ApJ...679.1338R} that are consistent with the CMU model using a centrifugal radius of $170\,$AU. 
The blue and red-shifted flows were found to have lengths greater than $1200\,$AU with  
specific angular momentum of $2.3-2.9\times10^{-3}$\jsunit\ and a mass accretion rate onto the disk of $1.2\times10^{-6}$\mdotunit. 
Lastly, in IRAS 03292+3039, \citet[][]{Pineda2020NatAs.tmp..151P} find an asymmetric streamer in HC$_{3}$N$\,$($J=10\rightarrow9$) with a length of $\sim10,500\,$AU that matches the CMU model in the inner part, and seems to have less velocity than expected by the model in the outer part. 
They estimate a mass-infall rate of the streamer to be $\sim10^{-6}$\mdotunit, comparable to the other observed accretion flows.
They also consider two different accretion scenarios, where the streamers seen are either part of the parental dense core or coming from beyond the parent dense core. 
They postulate that because the molecules detected in the accretion flow are ``chemically fresh", that the streams are coming from the outside the dense core, where they do not have a chance to be depleted.
While we have a few examples of these infalling accretion flows, many others report extended structures around other sources, but they are often disregarded as being part of the outflow cavity, due to their proximity to the observed molecular outflows \citep{Yen2017ApJ...834..178Y,Hull2020ApJ...892..152H, Gouellec2019ApJ...885..106L}.
Since everything is projected onto the plane-of-sky, it is not immediately apparent whether these emission are truly connected to the outflow or actually infalling material. 

\begin{deluxetable*}{cccccc}[t!]
\tablewidth{\textwidth}
\tablecaption{Summary of Image Results}
\label{tab:image}
\tablehead{
& & \multicolumn{2}{c}{This Work} & \multicolumn{2}{c}{Yen et al. 2017} \\
\cmidrule(lr){3-4}\cmidrule(lr){5-6}
Line/Continuum & \colhead{Rest Frequency} & \colhead{Beam Size (PA)} & \colhead{$\sigma_{\mathrm{rms}}$} & \colhead{Beam Size (PA)} & \colhead{$\sigma_{\mathrm{rms}}$}
}
\startdata 
\COan & 219.56035$\,$GHz & $1.10\arcsec\times1.04\arcsec\,(86.81^{\circ})$ & $3.47$ & $0.53\arcsec\times0.48\arcsec\,(21^{\circ})$ & 2.9\\
\SOn & 219.94944$\,$GHz & $1.10\arcsec\times1.04\arcsec\,(84.48^{\circ})$ & $4.20$ & $0.53\arcsec\times0.47\arcsec\,(20^{\circ})$ & 3.5\\
\COcn & 230.53800$\,$GHz & $1.15\arcsec\times1.04\arcsec\,(-83.65^{\circ})$ & $4.50$ & $0.51\arcsec\times0.46\arcsec\,(24^{\circ})$ & 2.5\\
\continuum & 234.00000$\,$GHz & $1.10\arcsec\times1.04\arcsec\,(-86.79^{\circ})$ & $1.89$ & $0.49\arcsec\times0.46\arcsec\,(8^{\circ})$ & 0.2\\
\enddata
\tablecomments{Rest frequencies of the molecular lines are taken from the \textit{Splatalogue} online database (\url{https://splatalogue.online//}). The noise level ($\sigma_{\mathrm{rms}}$) was calculated using channels 5-15 in each image cube and in an off-center region in the continuum image. The noise for the molecular line images is in units of mJy$\,$beam$^{-1}\,$channel$^{-1}$, while the noise for the continuum image is in units of mJy$\,$beam$^{-1}$. 
} 
\end{deluxetable*}

In this paper, we seek to explore this issue by using ALMA archival data of CO isotopologues (\COa, \COb\ and \COc) along with \SO\ to study the velocity components of extended emission around a young Class 0 source with a confirmed Keplerian disks, Lupus 3-MMS (IRAS 16059-3857) and compare with a simple analytical infall model (CMU Model, see Section \ref{sec:cmu} for a complete description). 
Lupus 3-MMS is a Class 0 protostar in the Lupus III molecular cloud with $T_{\mathrm{bol}}=39\,$K and $L_{\mathrm{bol}}=0.41\,L_{\odot}$ \citep{Dunham2013AJ....145...94D}.
\citet{Yen2017ApJ...834..178Y} find a Keplerian disk radius of $130\,$AU around a protostellar mass of $0.3\,M_{\odot}$ from kinematic modeling of \COa.
They also mention seeing extensions in \COa\ with lengths of 2\arcsec–4\arcsec\ along the cavity wall of the outflow, though they do not analyze these structures. 
The kinematic evidence of a rotationally supported disk, the available archival observations of molecular lines used to identify accretion flows in other sources and some hints at possible extended structures make this a promising source for our purpose.

The outline of this paper is as follows. Section \ref{sec:observations} describes the archival data used along with the calibration and image cleaning process. 
Section \ref{sec:results} shows the resulting image maps and identifies the different extended structures around our source. 
Section \ref{sec:analysis} compares the velocity of the extended structures we found to the outflow velocity structure and also to a simple infall model. 
Section \ref{sec:discussion} discusses the analysis of our model, previously observed accretion flows around other sources and MHD simulations. 
Section \ref{sec:conclusion} summarizes the paper and our main conclusions. 


\begin{figure*}[ht!]
\centering
\includegraphics[trim=0cm 0cm 0cm 0cm, clip, width=\textwidth]{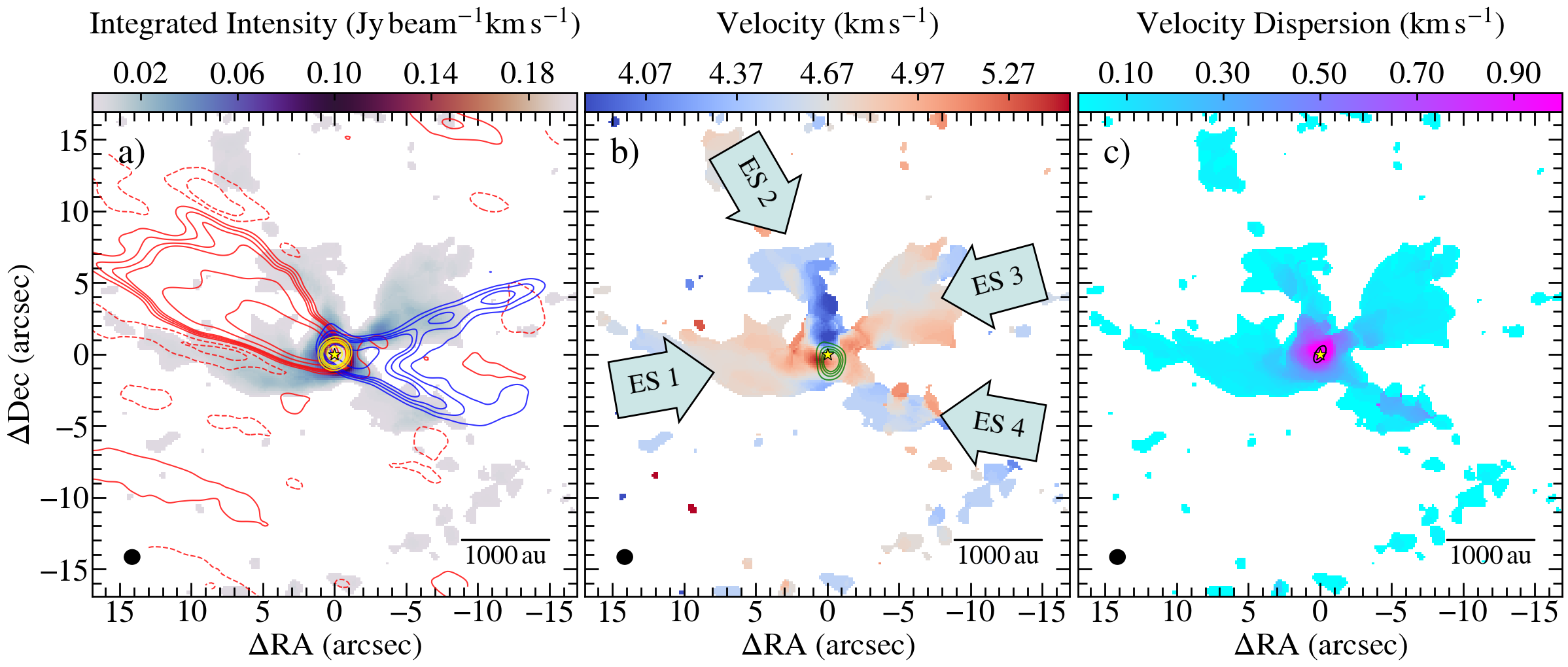}
\caption {\COan\ integrated intensity (\textbf{a}), intensity-weighted velocity (\textbf{b}) and intensity-weighted velocity dispersion (\textbf{c}) maps. 
Red and blue contours represent the red and blue-shifted outflows in \COc, respectively, at levels of -10, -5, 5, 10, 15, 20, 50 and 100$\sigma$, where $\sigma=22.35\ \mathrm{mJy\,beam^{-1}\,km\,s^{-1}}$. 
Yellow contours in the center show the \continuum\ emission at the same levels as \COc, where $\sigma=1.89\ \mathrm{mJy\,beam^{-1}}$. 
Green contours represent the \SO\ emission at the same levels as \COc, where $\sigma=2.40\ \mathrm{mJy\,beam^{-1}\,km\,s^{-1}}$. 
The yellow star denotes the position of Lupus 3-MMS. 
The arrows point to four extended structures (ES) of interest.
All maps were made by excluding pixels under $5\sigma$ from the image cube, where $\sigma=3.47\, \mathrm{mJy\,beam^{-1}\,channel^{-1}}$.}
\label{fig:c18o_moments}
\end{figure*}

\section{Observations} \label{sec:observations}
We use previously observed spectral line data of Lupus 3-MMS from the ALMA archive (Project ID: 2013.1.00879.S, PI: Hsi-Wei Yen). 
The baseline lengths at 220$\,$GHz range from 18$\,$m to 646$\,$m (13$\,$k$\lambda$ to 498$\,$k$\lambda$).
This corresponds to a maximum recoverable scale of $\sim6.2\arcsec$ and $\sim5.9\arcsec$, respectively.
The pointing center was RA$\,\mathrm{(J2000)}=16^{\mathrm{h}}09^{\mathrm{m}}18^{\mathrm{s}}\!.07$ and Dec$\,\mathrm{(J2000)}=-39^{\mathrm{h}}04^{\mathrm{m}}51^{\mathrm{s}}\!.6$. 
We focus on three of the observed spectral windows centered on different molecular lines that show detections: \COan, \COcn and \SOn. 
The calibration was done by using the  Common Astronomy Software Applications (CASA; v5.1.2+4.6.2) to run the calibration script provided with the data \citep{CASA}.
Self-calibration was not applied.
The calibrated molecular line data have channel widths of $61\,$kHz.
We performed continuum subtraction of the molecular line data using the continuum spectral window and the line-free channels of the molecular line spectral windows using the CASA task \textit{``uvcontsub''}.  
The data were cleaned using the CASA task \textit{``tclean''} with natural weighting to give slightly lower resolution, but have the best signal-to-noise ratio after performing the Fourier-transform.
We used the default ``hogbom'' deconvolver and apply a UV taper of 100$\,$k$\lambda$.
This taper weights the visibilities to shorter baselines ($<$100$\,$k$\lambda$) and is done in order to try to reveal more extended structure around the source. 
We tested other UV tapers and found that tapers $>$100$\,$k$\lambda$ resolved out most of the extended emission and tapers <100$\,$k$\lambda$ became too noisy due to the sensitivity decreasing from down-weighting the data.
We produce four molecular line image cubes, each with an image size of 320 pixels, a cellsize of 0.15\arcsec\ per pixel, and a velocity resolution of 0.1$\,$km$\,$s$^{-1}$ and one continuum image, also with an image size of 320 pixels and a cellsize of 0.15\arcsec\ per pixel. 


\section{Results} \label{sec:results}
A summary of our image results are shown in Table \ref{tab:image}. 
The use of natural weighting with a UV taper produces images with a beam size of $\sim 1^{\prime\prime}$. 
The shorter baselines in our data contain the flux from extended emission, which is crucial for our analysis. 
We compare to the image results obtained by \citet{Yen2017ApJ...834..178Y}, who cleaned the data using Briggs weighting and a robust parameter of $+0.5$ to achieve beam sizes of $\sim 0.5^{\prime\prime}$. 
Their higher resolution images are better suited toward resolving the Keplerian disk and analyzing its properties. 
As a note, we only show the \COc, \SO\ and \continuum\ results as contours, since our focus is mainly on the potential accretion flow tracer \COa; we will briefly describe their results in the order as they appear in the figures. 

Figure \ref{fig:c18o_moments} shows our \COa\ results of Lupus 3-MMS. 
We uncover four clear structures in \COa\ that extend out from the central region and follow the edges of the \COc\ outflow (Figure \ref{fig:c18o_moments}a). 
The structures appear to extend much further than previously noted by \citet{Yen2017ApJ...834..178Y}. 
Each is on the order of a few thousand au in length. 
The bipolar outflow has two clear red and blue-shifted components, shown as red and blue-shifted contours (Figure \ref{fig:c18o_moments}a).
The red-shifted outflow appears to be straight, while the blue-shifted component is slightly bent towards the north direction.
The outflows were previously found to have position angles (PA) of $60^{\circ},\ 265^{\circ}$ and inclination angles ($i$) of $65^{\circ}$ and $50^{\circ}$ for the red and blue-shifted components, respectively.
The \continuum\ emission in yellow contours shows a single compact structure with a central position of $\mathrm{RA\,(J2000)} = 16\mathrm{h}09\mathrm{m}18.09\mathrm{s}$ and $\mathrm{Dec\,(J2000)} = -39\mathrm{d}04\mathrm{m}53.3\mathrm{s}$. 
There is a clear velocity gradient in the central region showing the rotationally supported disk (Figure \ref{fig:c18o_moments}b). 
The \SO\ emission in green contours also shows a compact structure slightly off-center from the source.
We label our four extended structures (ES) with arrows.
ES1 is a structure extending $\sim2000$\au\ under the red-shifted outflow. 
The emission from ES1 is mostly red-shifted in velocity and attaches to the red-shifted part of the disk. 
Near the far end of the flow, the emission becomes slightly blue-shifted.
ES2 is a structure extending $\sim1300$\au\ above the red-shifted outflow.
The emission from ES2, unlike the outflow, is blue-shifted and has a higher velocity closer to the disk. 
ES3 is a structure extending $\sim1800$\au\ above the blue-shifted outflow.
The emission from ES3 seems to have a mixture of blue and red-shifted velocities.
ES4 is a structure extending $\sim1600$\au\ below the blue-shifted outflow. 
The emission from ES4 also has a mixture of blue and red-shifted velocities, with the outer part being more blue and the inner part turning to red.
The velocity dispersion in the central region is high due to the faster Keplerian rotation of the disk.
All of the extended structures have a relatively low level of velocity dispersion ($<0.5\,\mathrm{km\,s^{-1}}$), indicating that the gas is not being perturbed (Figure \ref{fig:c18o_moments}c). 
Except for the right edge of the red-shifted outflow, we found the velocity dispersion to be $>1.0\,\mathrm{km\,s^{-1}}$ in the \COc\ emission. 
This hints that the gas may not be affected by the outflow and could actually be a separate component, rather than just a part of the outflow cavity. 


\begin{figure*}[t!]
\centering
\includegraphics[trim=0cm 0cm 0cm 0cm, clip, width=\textwidth]{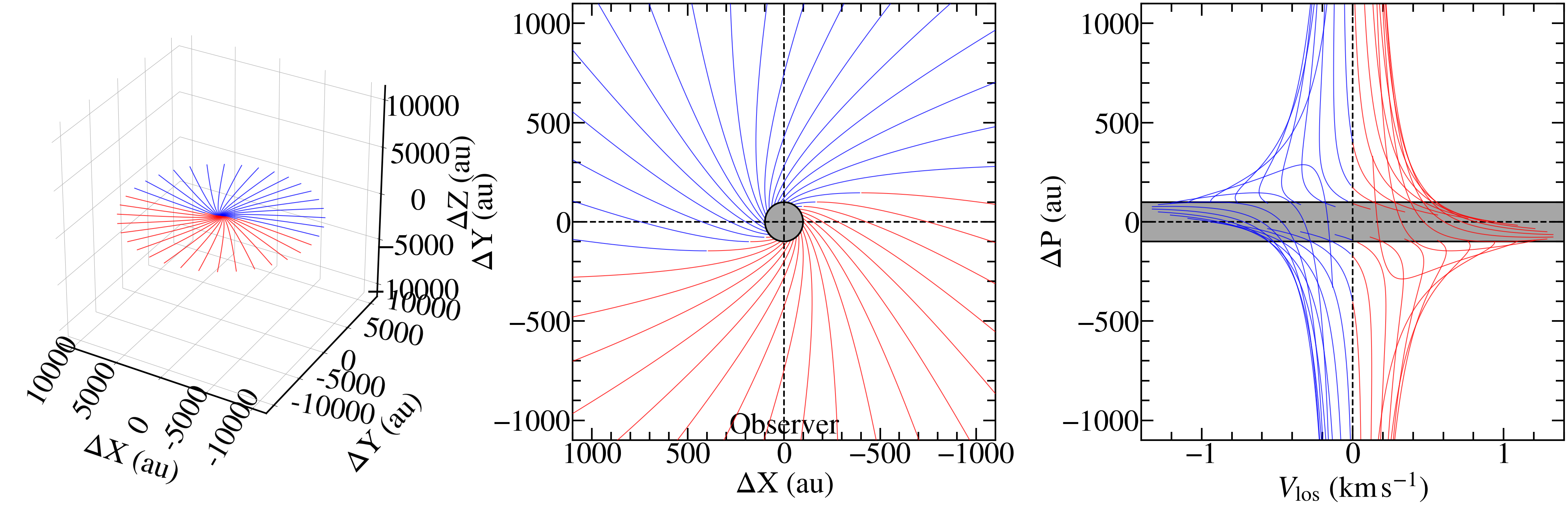}
\caption {CMU model plots in 3D space (left), viewing the inner region from the top-down (center) and the position-velocity structure in the inner region (right). 
In this example, we use $M_{*}=0.1\,\mathrm{M_{\odot}}$, $r_c=100\,\mathrm{au}$, $R_0=10000\,\mathrm{au}$ with $\mathrm{PA}=90^{\circ}$ and $i=90^{\circ}$ for an edge-on disk configuration. 
For visualization purposes, we only plot streamlines at $\theta_0=90^{\circ}$ (in the disk plane).
The streamlines are blue or red depending on when the line-of-sight velocity ($v_y$ in our model) is negative or positive. 
The gray areas mark the area within the centrifugal (disk) radius. 
The direction of the observer is also marked in the center plot.
See \citet{Tobin2012ApJ...748...16T} for additional plots and information.}
\label{fig:cmu_test}
\end{figure*}

\section{Analysis} \label{sec:analysis}
\subsection{The CMU Model}\label{sec:cmu}
The CMU (Cassen-Moosman-Ulrich) model describes the infalling particle trajectories (or streamlines) around a point mass in the gravitational collapse of a rotating, spherically symmetric cloud  \citep{Ulrich1976ApJ...210..377U,Cassen1981Icar...48..353C,Chevalier1983ApJ...268..753C,Terebey1984ApJ...286..529T}. 
The assumptions of this model are: 
\begin{enumerate}
  \item The cloud is undergoing solid-body rotation at a uniform angular velocity $(\Omega)$  
  \item The central star is treated as a point mass, meaning self-gravity of the envelope is not considered
  \item The particle starts from a spherical surface at a radius $(r_0)$ and follows a parabolic trajectory until it collides with the disk
  \item Specific angular momentum along the trajectory is conserved
  \item Pressure forces are negligible, meaning trajectories are ballistic (under the constraint that streamlines do not intersect)
  \item Disk and envelope mass are not considered
  \item Magnetic fields are not considered
\end{enumerate}
Parabolic trajectories of the infalling matter are described by the trajectory equation:

\begin{equation} \label{eq:1}
r=\frac{j^2}{GM_{*}}\frac{1}{1-\cos\nu}
\vspace{0.2cm}
\end{equation}
where $j$ is the specific angular momentum, $M_{*}$ is the stellar mass and $\nu$ is the direction angle of the particle measured from the origin to apastron. 
Due to the spherical geometry of this model, we have the following angular relations:

\begin{equation} \label{eq:2}
\cos\theta=\cos\nu\cos\theta_0
\end{equation}

\begin{equation} \label{eq:3}
\tan\nu=\tan(\phi-\phi_0)\sin\theta_0
\vspace{0.2cm}
\end{equation}
where $\theta_0$ is the polar angle of the orbital plane for the particle trajectory, $\phi_0$ is the azimuthal angle of the orbital apastron, $\theta$ and $\phi$ are the polar and azimuthal angles of the particle after a time $t$.
The radius of the point at which the streamlines intersect with the disk ($r_p$; i.e., the semi-latus rectum of the parabolic trajectory) depends on the initial polar angle $\theta_0$ and is described by 

\begin{equation} \label{eq:4}
r_p=\frac{r_0^4 \Omega^2}{GM_{*}}\sin^2 \theta_0
\vspace{0.2cm}
\end{equation}
This equation is a result of angular momentum conservation. 
When $\theta_0=90^{\circ}$, the radius at which the particle collides with the disk is maximum. 
This is refered to as the centrifugal radius ($r_c$) and is described by

\begin{equation} \label{eq:5}
r_c=\frac{r_0^4 \Omega^2}{GM_{*}}
\vspace{0.2cm}
\end{equation}
In a spherical rotating system, the specific angular momentum is expressed as 

\begin{equation} \label{eq:6}
j^2=r_0^4 \Omega^2\sin^2\theta_0
\end{equation}
Using Equation \ref{eq:2} and combining Equations \ref{eq:5} and \ref{eq:6}, we can rewrite the trajectory equation as 

\begin{equation} \label{eq:7}
\frac{r}{r_c}=\frac{\sin^2\theta_0}{1-\cos\theta/\cos\theta_0}
\vspace{0.2cm}
\end{equation}
Equations \ref{eq:2}, \ref{eq:3} and \ref{eq:7} lay the framework for the position of the particles in the CMU model. 
To complete the framework, we also need the velocity of the particles. 
In spherical coordinates, the CMU model describes the velocities as 

\begin{equation} \label{eq:8}
v_r=-\left(\frac{GM_*}{r}\right)^{1/2}\left(1+\frac{\cos\theta}{\cos\theta_0}\right)^{1/2}
\end{equation}

\begin{equation} \label{eq:9}
v_\theta=\left(\frac{GM_*}{r}\right)^{1/2}\left(\frac{\cos\theta_0-\cos\theta}{\sin\theta}\right)\left(1+\frac{\cos\theta}{\cos\theta_0}\right)^{1/2}
\end{equation}

\begin{equation} \label{eq:10}
v_\phi=\left(\frac{GM_*}{r}\right)^{1/2}\left(\frac{\sin\theta_0}{\sin\theta}\right)\left(1-\frac{\cos\theta}{\cos\theta_0}\right)^{1/2}
\vspace{0.2cm}
\end{equation}
In order to quantitatively compare to observations, the spherical geometry is easily converted to a Cartesian system where it can then be rotated using rotation matrices to match the position angle (PA) and inclination angle ($i$) of the source. 
Figure \ref{fig:cmu_test} displays the CMU model plots. 
We are able to construct a 3D representation of the model, as well as make projections onto various axes and produce position-velocity (PV) diagrams. We take $x$ and $z$ to be the RA and Dec axes, and take $v_y$ to be the line-of-sight velocity ($v_\mathrm{los}$). The observed velocity ($v_\mathrm{obs}$), which we can use to directly compare with our data, is then $v_\mathrm{sys}+v_\mathrm{los}$, where $v_\mathrm{sys}$ is the system velocity.

\begin{figure*}[ht!]
\centering
\includegraphics[trim=0cm 0cm 0cm 0cm, clip, width=\textwidth]{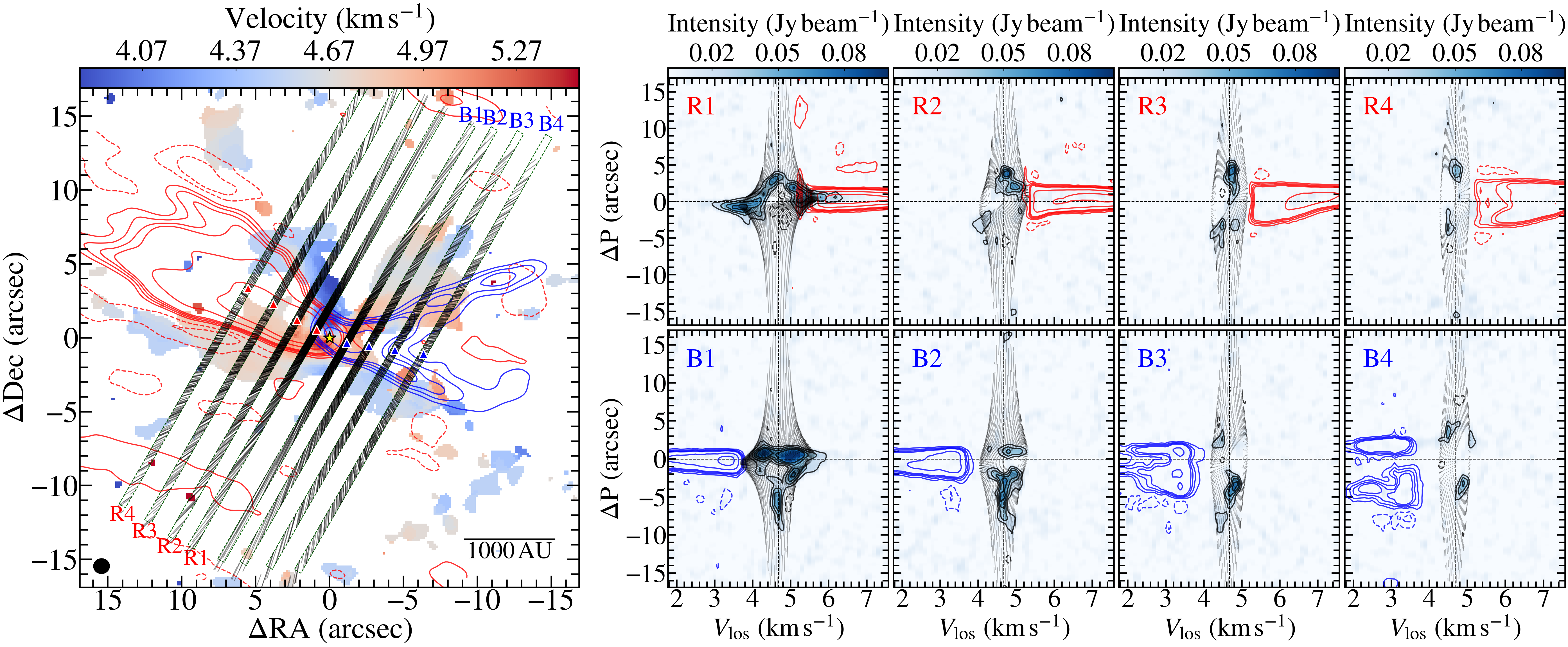}
\caption {PV cuts using the disk PA taken at different points along each outflow axis. The positions of each cut are overlaid on the \COa\ intensity-weighted velocity map and the \COc\ integrated-intensity contours from Figure \ref{fig:c18o_moments} (left). We plot the CMU model trajectories inside each cut (black lines) using $\theta_0$ from $30^{\circ}-150^{\circ}$ and $\phi_0$ from $0^{\circ}-360^{\circ}$, both in steps of $1^{\circ}$. The colored triangles show the center of each cut. The PV diagrams for each cut (right) of \COa\ (background and black contours) and \COc\ (red and blue contours) show the cuts along the red-shifted (right, top row) and blue-shifted (right-bottom) outflows, with distance increasing to the right. All contours show levels of -10, -5, 5, 10, 15, 20, 50 and 100$\sigma$, where $\sigma($\COa$)=3.37, 3.11, 3.03, 3.12, 3.13, 2.98, 3.39, 3.16\,$mJy$\,$beam$^{-1}$ and $\sigma($\COc$)=12.07, 13.65, 20.69, 21.02, 6.80, 7.34, 6.79, 6.93\,$mJy$\,$beam$^{-1}$, from left to right, top to bottom. The noise values were calculated from a square region in the bottom left corner of the PV image with no emission.}
\label{fig:c18o_outflow_compare}
\end{figure*}

\begin{figure}[ht!]
\centering
\includegraphics[trim=0cm 0cm 0cm -2cm, clip, width=\columnwidth]{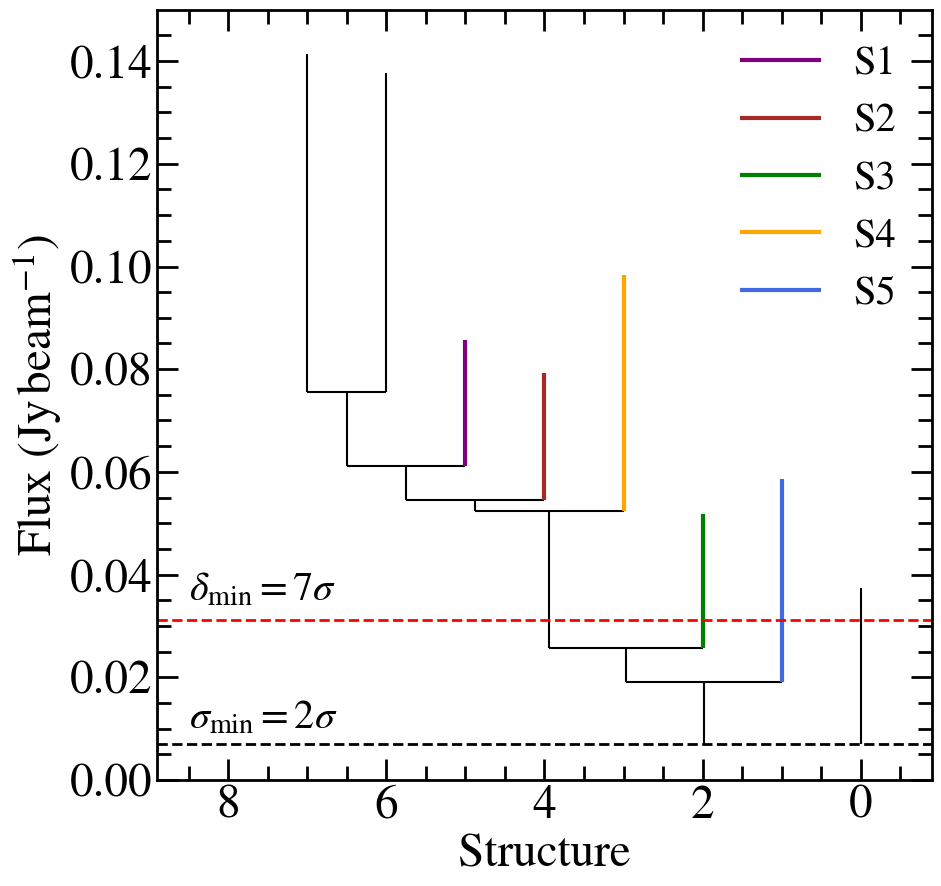}
\caption {Dendrogram tree of our \COa\ image cube. We label five leaf structures corresponding with our extended structures of interest, each labeled with a different color. We plot two horizontal lines for the minimum noise threshold ($\sigma_{\mathrm{min}}$) and the minimum significance for structures ($\delta_{\mathrm{min}}$), where $\sigma=3.47\,\mathrm{mJy\,beam^{-1}}$.}
\label{fig:dendro_tree}
\end{figure}

The CMU model describes trajectories infalling from everywhere in a globalized collapse scenario.
It is possible that asymmetric density structures in the envelope cause over-densities in the infalling material that resemble stream like structures \citep{Tobin2012ApJ...748...16T}. 
In this scenario, material may still be collapsing from everywhere around the disk, but only the over-dense regions are visible in observations. 
We explore this idea more in Appendix \ref{sec:streams}, where we test if various sized patches of infalling particles can produce accretion streams.
We find that streams can be produced in the different scenarios we tested. 
This supports the CMU model as a viable method to compare to accretion streams in observations.


\subsection{Disk Properties and System Velocity}\label{sec:disk}
Kinematic analysis is required to understand the nature of these extended structures. 
We need to ensure that the disk properties and system velocity are correct in order to effectively compare to our CMU model. 
Recently, \citet{Dzib2018ApJ...867..151D} found a new distance to the Lupus III molecular cloud, closer than that used by \citet{Yen2017ApJ...834..178Y}. 
Using \textit{Gaia} DR2 data, they find a new distance of $162\pm3\,$pc.
We adopt this new distance and scale the disk radius ($r_d$) and protostellar mass ($M_*$) previously derived from Keplerian rotation fitting. 
We use the original values of $r_d=130\,$au and $M_*=0.3\,$M$_{\odot}$ to find scaled values of $r^{\mathrm{scaled}}_d=105\,$au and $M^{\mathrm{scaled}}_*=0.24\,$M$_{\odot}$ at this new distance.
In our model, we assume the value for $r^{\mathrm{scaled}}_d$ is equal to the centrifugal radius, $r_c$.
Additionally, since our cubes were made with a slightly different velocity resolution and weighting than that used by \citet{Yen2017ApJ...834..178Y}, we refit the center of the \COa\ spectrum with a double peaked gaussian to find a new centroid velocity ($v_{\mathrm{sys}}$) for comparing our model with observations. 
We chose a circular region with a radius of  $\sim1^{\prime\prime}$ and extracted the spectrum from the image cube. 
Our fitting results give a new value of $v_{\mathrm{sys}}=4.67\pm0.59\,$km$\,$s$^{-1}$.
A correct value for $v_{\mathrm{sys}}$ is crucial for getting the correct observed velocity, though the value derived from kinematic fitting of the disk by \citet{Yen2017ApJ...834..178Y} is within error.

\subsection{Outflow Velocity Structure}\label{sec:outflow}
We do a number of comparisons between the \COa\ emission that traces a number of extended features and the \COc\ emission that traces the outflow. 
We first make PV cuts along the disk (PA=$150^{\circ}$) and outflow (PA=$60^{\circ}$) axis with two widths (0.5\arcsec\ and 20.0\arcsec) to examine the velocity of the structures compared to the outflow. 
The PV cuts along the disk axis show that \COc\ extends to much higher velocities, with the structures becoming much different in the wider 20.0\arcsec\ cut. 
The PV cuts along the outflow axis show very different velocity structures in both cuts, with the \COc\ exhibiting a parabolic shape moving away from the center and increasing in velocity, while the \COa\ emission is very centralized.
These results are described in more detail and shown in Appendix \ref{sec:pvcomp}.

\begin{figure*}[ht!]
\centering
\includegraphics[trim=0cm 0cm 0cm 0cm, clip, width=\textwidth]{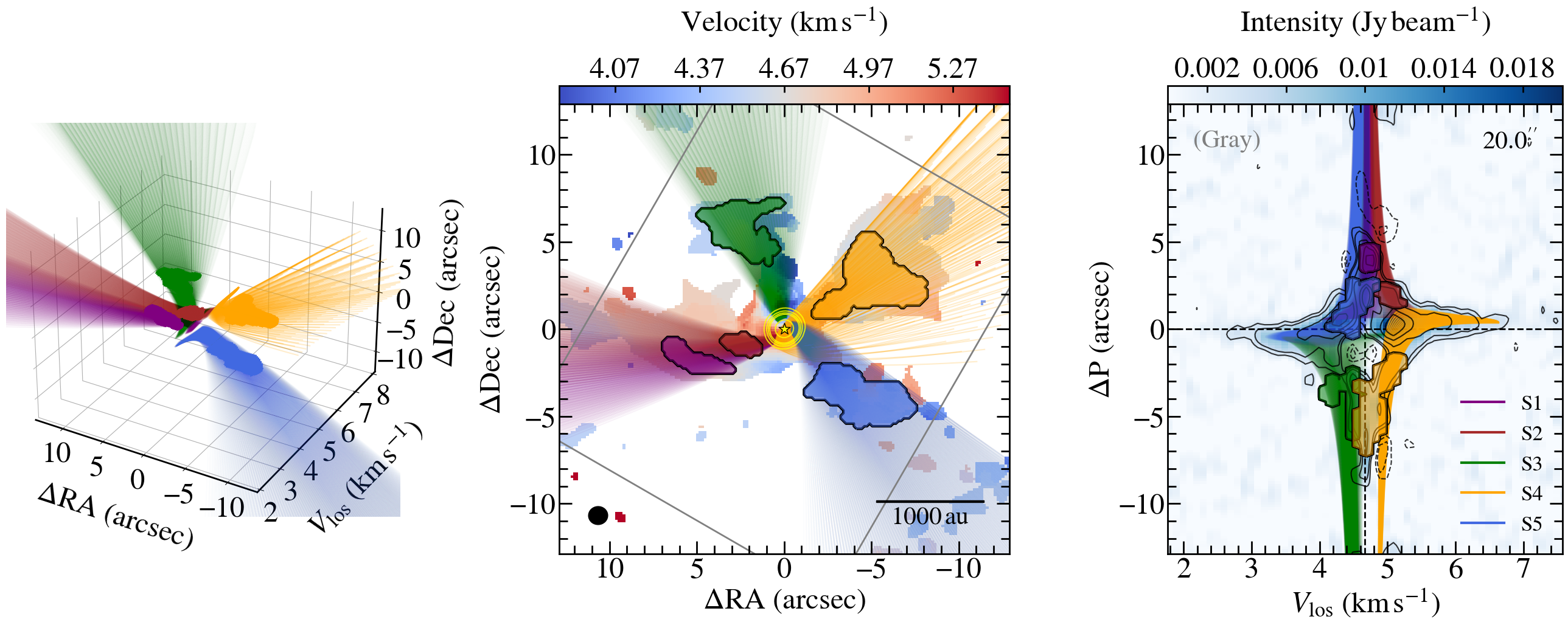}
\caption {Position-Position-Velocity (PPV) matching of the CMU model to the Dendrogram structures in three different views: PPV space (left), PP space (center) and PV space (right). The colors all correspond to the same structures found in Figure \ref{fig:dendro_tree}. The dendrogram structures are all shown as surfaces, while the trajectories from the CMU model are shown as streamlines. The gray box (center figure) shows the PV cut along the disk plane with a width of 20.0\arcsec. The \COa\ contours in the PV diagram are show levels of -10, -5, -3, 3, 5, 10, 15, 20, 50 and 100$\sigma$, where $\sigma=0.91\,\mathrm{mJy\,beam^{-1}}$.}
\label{fig:c18o_dendrogram}
\end{figure*}

Instead, we make cuts along the disk axis at different points along the red and blue-shifted outflows as another way to disentangle the outflow from the extended emission.
The position angle (PA) of the disk previously found by \citet{Yen2017ApJ...834..178Y} is 150$^{\circ}$.
Figure \ref{fig:c18o_outflow_compare} shows the positions of each cut overlaid on the \COa\ intensity-weighted velocity map (Figure \ref{fig:c18o_outflow_compare}, left) and each of their respective PV diagrams (Figure \ref{fig:c18o_outflow_compare}, right).
For each PV cut, we plot the CMU model curves corresponding to a globalized collapse scenario with $\theta_0$ from $30^{\circ}-150^{\circ}$ and $\phi_0$ from $0^{\circ}-360^{\circ}$, both in steps of $1^{\circ}$, around a protostellar mass ($M_*$) of 0.24$\,$M$_{\odot}$ with the centrifugal radius ($r_c$) set to be $105\,$au. 
We find the \COa\ emission structure in the PV diagrams to be fairly consistent with the overall structure of the CMU model.
The \COa\ emission lies mostly within the extent of the model curves. 
As well, we find some gaps in the CMU model  trajectories seen in the PV diagrams.
The \COa\ emission mostly avoids these gaps while staying within the extent of the model.
The high-velocity \COa\ only appears around the center.
In contrast, the \COc\ emission structure is much higher velocity in each of the PV cuts, and seem to become more circular in shape and increase in velocity as we move further away along the outflow axis. 
The two molecules are most certainly tracing different dynamics in this system. 

\begin{deluxetable*}{lccccccc}[ht!]
\tablewidth{\textwidth}
\tablecaption{Dendrogram Matching}
\label{tab:matching}
\tablehead{\vspace{-0.1cm}
Dendrogram & \colhead{$r_0$} & \colhead{$\theta_0$ range} & \colhead{$\phi_0$ range} & \colhead{\# of Matching} & \colhead{\# of Dendrogram} & \colhead{Fit}\\
Structure & \colhead{(au)} & \colhead{($^{\circ}$)} & \colhead{($^{\circ}$)} &  \colhead{Points} & \colhead{Points} & \colhead{Percentage} 
}
\startdata 
S1 & 10000 & 118--139 & 120--142 & 279  & 279  & 100\% \\ 
S2 & 10000 & 111--128 & 90--116 & 157  & 157  & 100\% \\ 
S3 & 10000 & 125--157 & 298--368 & 454  & 469  & 96.80\% \\ 
S4 & 10000 & 50--58  & 52--77  & 359 & 1199  & 29.94\% \\ 
S5 & 10000 & 31--74  & 201--252  & 1050 & 1060 & 99.06\% \\ 
\enddata
\tablecomments{The step size for both $\theta_0$ and $\phi_0$ is set to 1$^\circ$. The fitting percentage is calculated from the number of matching points divided by the total number of points in the dendrogram.}
\end{deluxetable*}

\subsection{Infall Velocity Structure} \label{sec:infall}

Since the emission of the \COa\ emission shows the most disconnect with the \COc\ outflow emission, we begin by investigating these structures more directly and compare them to the CMU infall model. 
To do this, we use the same method as in \citet{CheongThesis2018}, who apply a ``dendrogram'' algorithm \citep{Rosolowsky2008ApJ...679.1338R} to their data to pick out structures from their \COa\ image cube in Position-Potision-Velocity (PPV) space. 
The dendrogram algorithm works by using the brightest pixels in the image cube to construct a tree, then adding fainter and fainter pixels until a new local maximum is found and a new structure is created. These structures form the ``leaves'' of the tree and are then connected by ``branches'', which is a pixel that is not a local maximum. 
All of these structures eventually merge and create a tree. 
We input a number of parameters into the algorithm, including a minimum noise threshold for the tree ($\sigma_{\mathrm{min}}$), a minimum significance for structures ($\delta_{\mathrm{min}}$) and the minimum number of pixels that a structure should contain in order to remain an independent structure ($n_{\mathrm{pix,min}}$). 
We set $\sigma_{\mathrm{min}}$ to be $2\sigma$, $\delta_{\mathrm{min}}$ to be $7\sigma$ and $n_{\mathrm{pix,min}}$ to be 150 pixels. 
These parameters were decided by testing different values to find the set that showed the most coherent structures.
We run the algorithm on our \COa\ image cube and find a total of 8 structures.
Figure \ref{fig:dendro_tree} shows the Dendrogram tree structure of our result.
Out of these 8, only 5 are correlated with the extended structures of interest we previously labeled in Figure \ref{fig:c18o_moments}.

\begin{figure*}[ht!]
\centering
\includegraphics[trim=0cm 0cm 0cm 0cm, clip, width=\textwidth]{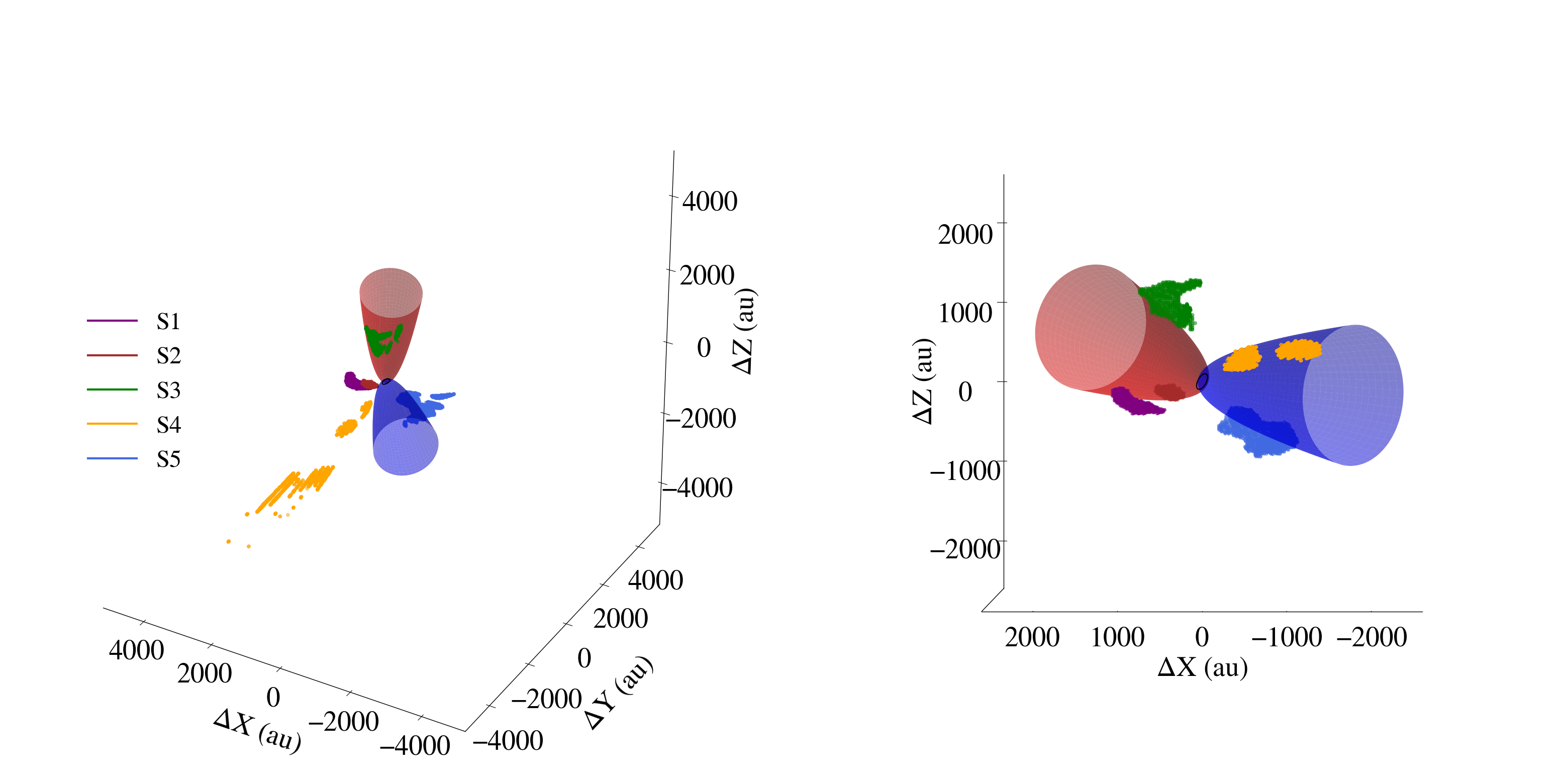}
\caption {Position-Position-Pelocity (PPP) trajectories using the matched points from the CMU model. The x and z axes correspond to the RA and Dec, while the y-axis is the line-of-sight axis. The colored dots all correspond to the same colored structures found in Figure \ref{fig:dendro_tree}. The red and blue cones represent the red and blue-shifted outflows using the position and inclination angles derived by \citet{Yen2017ApJ...834..178Y}. An animation of this figure is available where it is rotated about the z-axis. The sequence repeats 3 times and the real-time duration of the animation is 42 seconds.}
\label{fig:cmu_3d}
\end{figure*}

We explore these structures in position-position-velocity (PPV) space and compare their morphologies to the CMU model (Figure \ref{fig:c18o_dendrogram}). 
Figure \ref{fig:c18o_dendrogram} (left) shows the 3D structures of our dendrograms and CMU model streamlines in PPV space. 
Figure \ref{fig:c18o_dendrogram} (center, right) decomposes the PPV space into position-position (PP) and position-velocity (PV) spaces.
We start by going through each structure and finding a CMU streamline that matches in both position and velocity to a pixel in the dendrogram structure. 
We set criteria for a point along the CMU trajectory to match with a point in the PPV dendrogram structure.
The point along the streamline must lie within the RA and Dec of the dendrogram point $\pm$ half of the image cube pixel width and the line-of-sight velocity $\pm$ half of the image cube velocity resolution. 
Our pixel size of 0.15\arcsec corresponds to $\sim24$\au, and the velocity resolution was set to 0.1$\,\mathrm{km\,s^{-1}}$.
Each streamline has an initial radius ($r_0$) of 10,000\au\ and ends at the centrifugal radius, which we assume is equal to the disk radius of 105\au. 
The mass of the central point source is taken to be 0.24\msun, i.e. the mass of the central protostar. 
We search for a streamline that has at least one matching point in each structure, then increase the range of the trajectory parameters ($\theta_0$ and $\phi_0$) angles until the number of matching points reaches a maximum. 
We use the maximum number of matching points divided by the number of points in the dendrogram to calculate a fitting percentage.
This fitting percentage is to quantify how many pixels within the structure can be matched with the CMU model. 
We find most of our streams are consistent with the CMU model (Table \ref{tab:matching}). 
S1 and S2 match 100\%, while S3 and S5 being $>95$\%.
S4 shows a large discrepancy with a fit percentage of $<30$\%, due to very low velocity points (near the system velocity) seen in the PV diagram (Figure \ref{fig:c18o_dendrogram} right) between -3\arcsec\ and -7\arcsec. 

Using the matching points found from the CMU model ($x$, $z$ and $v_y$), we can construct a representation in 3D position-position-position (PPP) space by taking the $z$ coordinate at the same indices (Figure \ref{fig:cmu_3d}). 
Additionally, we plot the red and blue-shifted outflows using the parameters derived by \citet{Yen2017ApJ...834..178Y}.
They adopt an axisymmetric wind-driven outflow model \citep{Lee2000ApJ...542..925L} and fit the position and inclination angles of the outflows.
In PPP space, we see that the CMU points with a $>95$\% fit seem to be infalling more in line with the disk plane, while S4 aligns more next to the edge of the outflow and shows large gaps of missing points along the flow. 
The flows are all found at different positions around the disk.
S1 and S2 are both found close to each other, indicating that they may actually be part of the same infalling structure, but could not be identified as one in the dendrogram. 
S3 is falling from above the disk, while S5 is coming in from below. 
S4 is located almost completely along the line-of-sight and runs adjacent to the blue-shifted outflow.

We derive several properties of our accretion flows using the matched 3D components of our model (Table \ref{tab:properties}). 
The distances are roughly similar in 2D space, ranging between 591\au\ and 1416\au. 
The differences arise when the distance is calculated from the 3D coordinates.
The largest discrepancy of more than 7000\au\ in S4 and is caused by the matching points to be mainly along the line-of-sight. 
S5 is also slightly more along the line-of-sight, giving a discrepancy of $\sim520$\au.
S1, S2, and S3 are mostly in the plane-of-sky, and only change by values of $\sim100$\au\ or less. 
Next, we make \COa\ column density maps using the method from \citet{Mangum_2015} in order to estimate the mass of each flow. 
We describe and show the column density maps for each structure in Appendix \ref{sec:cdm}.
Summing the column density in each structure and assuming a \COa\ abundance with respect to H$_2$, [\COa/H$_2$], of $1.7\times 10^{-7}$ \citep{Frerking1982ApJ...262..590F}, we estimate the flow masses to be between $1.3-5.1\times 10^{-3}$\msun.
We kinematically derive the infall timescale based on the lengths and infalling velocities of the flows, finding a range of $1.5-9.8\times10^{4}\,$yr.
Using the flow mass and the infall timescale, we estimate the mass-infall rate of the flows to be $0.5-1.1\times10^{-6}$\mdotunit.
The specific angular momentum along a trajectory is conserved, but can vary based on $\theta_0$.
We combine Equations \ref{eq:4} and \ref{eq:5} to derive $j=\sqrt{GM_* r_c}\sin \theta_0$, and find values of $\sim10^{-4}$\jsunit. 
Lastly, we use the specific angular momentum and multiply it by the flow mass to calculate the total angular momentum of each flow, getting values ranging from $0.76-7.75\times10^{-6}$\Jsunit.

\begin{deluxetable*}{lcccccccc}[htb!]
\tablewidth{\textwidth}
\tablecaption{Accretion Flow Properties}
\label{tab:properties}
\tablehead{\vspace{-0.1cm}
Dendrogram & \colhead{$r_{\mathrm{max,2D}}$} & \colhead{$r_{\mathrm{max,3D}}$} & \colhead{$N$} & \colhead{$M_{\mathrm{flow}}$} & \colhead{$t_{\mathrm{infall}}$} & \colhead{$\dot{M}_{\mathrm{infall}}$} & \colhead{$j$} & \colhead{$J$}\\
Structure & \colhead{(au)} & \colhead{(au)} & \colhead{(cm$^{-2}$)} & \colhead{(M$_{\odot}\,$)} & \colhead{(yr)} & \colhead{(M$_{\odot}\,$yr$^{-1}$)} & \colhead{(\jsunit)} & \colhead{(\Jsunit)}
}
\startdata 
S1 & 1128  & 1130  & $2.38\times10^{13}$ & $3.20\times10^{-3}$ & 3.55$\times10^{3}$ & $9.01\times10^{-7}$ & 4.76-6.40$\times10^{-4}$ & 1.52-2.05$\times10^{-6}$\\ 
S2 & 591   & 610   & $1.24\times10^{13}$ & $1.67\times10^{-3}$ & 1.54$\times10^{3}$ & $1.06\times10^{-6}$ & 5.71-6.77$\times10^{-4}$ & 0.95-1.13$\times10^{-6}$\\ 
S3 & 1349  & 1460  & $2.01\times10^{13}$ & $2.69\times10^{-3}$ & 5.51$\times10^{3}$ & $4.88\times10^{-7}$ & 2.83-5.94$\times10^{-4}$ & 0.76-1.60$\times10^{-6}$\\ 
S4 & 1367  & 8480  & $9.39\times10^{13}$ & $1.26\times10^{-2}$ & 2.25$\times10^{5}$ & $5.60\times10^{-8}$ & 5.55-6.15$\times10^{-4}$ & 6.99-7.75$\times10^{-6}$\\ 
S5 & 1416  & 1940  & $3.80\times10^{13}$ & $5.09\times10^{-3}$ & 9.81$\times10^{3}$ & $5.19\times10^{-7}$ & 3.73-6.97$\times10^{-4}$ & 1.90-3.55$\times10^{-6}$\\ 
\enddata
\tablecomments{The maximum 2D radius ($r_{\mathrm{max,2D}}$) is calculated from the maximum X and Z model values matching with the dendrogram structure, while the maximum 3D radius ($r_{\mathrm{max,3D}}$) is calculated from the maximum X, Z and corresponding Y model values. The column density ($N$) totals are assuming \COa\ excitation temperatures of 30$\,$K. The mass of the flows ($M_\mathrm{flow}$) is the H$_{2}$ mass, while the infall timescale ($t_\mathrm{infall}$) is calculated kinematically based on the lengths and velocities of the flows. The mass-infall rate ($\dot{M}_{\mathrm{infall}}$) is calculated using mass of the flows and the infall timescale. The specific angular momentum ($j$) values are the min-max values corresponding to the model positions matched with the dendrogram structures with the total angular momentum ($J$) is found by multiplying by the flow mass.}
\end{deluxetable*}

\section{Discussion} \label{sec:discussion}

\subsection{Outflow Cavities or Accretion Flows?} \label{sec:OCorAF}
We aimed to disentangle whether the extended emission in Lupus 3-MMS is part of the outflow cavity or are accretion flows. 
Outflow cavities are thought to be formed by a wind driven from the circumstellar disk of a protostar that pushes gas towards the outer edges of the outflow. 
This leads to questions such as whether the material in the outflow cavity has a similar velocity structure with the outflow, whether material can be recycled from the outflow and accrete back onto the disk, or if large-scale material can fall along the outflow cavities. 
Comparing the velocity structures of the extended structures and the outflow velocity structure yields almost no similarities. 
First, the velocity dispersion in the extended structures is low compared to at the center (Figure \ref{fig:c18o_moments}c).
One might expect the velocity in the outflow cavities to be perturbed if they were being influenced by the outflow.
Furthermore, the velocity along the \COc\ outflows is much higher than what is seen in \COa. 
If the material in the extended structures is really being pushed by the outflow, then for a given momentum input, we expect the lower density outer part should show a higher velocity. 
We do not expect missing flux to affect this point, as any high velocity components should still be captured by our observations.
Comparing with the globalized CMU model shows the \COa\ emission to stay within the extent of the model and avoids the holes seen in the PV diagrams. 
This seems to indicate that the structures are infalling. Further comparisons with the CMU model by isolating structures in PPV space using a dendrogram algorithm to confirm that these structures are infalling accretion flows.
S1, S2, S3 and S5 match the CMU model in PPV space with >95\% fits. 
S1 and S2 show remarkable similarity with fittings of 100\%.
It could be that these two are in fact part of the same structure, due to their proximity in position space, but are separated into two structures due to missing flux.
Follow-up observations with shorter baselines are needed to test this.
S3 and S5 have more than double the amount of pixels as the former two while still retaining a high fitting percentage of 96.80\% and 99.06\%, respectively. 
Thus, these structures in \COa\ can be better explained with an infall model and cannot be explained with an outflow model.
Another sign that accretion is taking place is the presence of a small patch of \SO\ near the red-shifted part of the disk (Figure \ref{fig:c18o_moments}b).
The presence of \SO\ is thought to be an indication of accretion shocks \citep{Sakai2014Natur.507...78S}. 
Since \SO\ only shows a blue-shifted component on one side of the disk, this could be an indication of asymmetric accretion.

\subsection{The Nature of S4} \label{sec:S4}
S4 has the lowest fitting percentage of less than 30\% from our CMU matching. 
S4 shows a much lower velocity component in the PV diagram (Figure \ref{fig:c18o_dendrogram}, right).
We explored was whether the material was further away from the protostar along the same trajectories. 
To test this, we increased the initial radius of the particle $r_0$ from 10,000\au\ (i.e., the typical dense core size of $\sim$0.1 pc) out to 20,000\au\ (i.e., twice the typical core size).
The CMU model traces the trajectory of the particle merely due to the gravitational pull of the central point mass.
Increasing the distance would allow for lower velocity components along the trajectory.
We re-ran the same fitting procedure as in Section \ref{sec:infall}, only to find the fitting percentage to increase from 29.94\% to 30.03\%, an increase of less than 1\%. 
Adjusting $\theta_0$ and $\phi_0$ also had no effect on this scenario. 
It is unlikely the unfit components are due to material that is further away along the line-of-sight. 
Additionally, material so far away runs into the issue of also being resolved out by the ALMA 12m array. 

We considered whether or not the material could be affected by magnetic braking. 
Recently, numerical simulations of protoplanetary disk formation by \citet{Lee2021A&A...648A.101L} found that in their simulations, material falling closer to the disk plane experienced more free-fall motion, while material falling closer to the outflow and the cavity was more prone to magnetic braking (see Figure 20 in their paper). 
S4 is located mainly along the line of sight towards the observer direction next to the blue-shifted outflow (Figure \ref{fig:cmu_3d}). 
If the low fitting percentage of S4 was due to magnetic braking, we might also expect S3 and S5 to have lower fitting percentages as well, as they show some components that are located positionally close to the outflow.
This scenario would need to be tested against MHD simulations in order to confirm this possibility, which is beyond the scope of this paper. 

The material along the streamline could be affected by the outflow. 
As we mentioned earlier, the velocity dispersion in each of the extended structures exhibits low, subsonic values. 
If the extended structures were being affected by the outflows, we would expect some kind of velocity perturbations due to the energetic out-flowing material. 
It is not likely that this stream is being affected by the outflow.

Missing flux at low velocities can bias the mean velocities away from the systemic velocity \citep{Yen2017AA...608A.134Y}. 
This could explain why there is a low velocity component that is not fit by the streamline model. 
S4 is found to go out to $\sim8000$\au\ in 3D space, making it longer than the other streams and more likely to have missing flux at these larger scales. 

Another quality of S4 is that it seems to have gaps in the fitting along the streamline. 
This could be due to the physical and chemical structure inside the envelope, causing different excitation conditions along the streamline. 
It may be possible that portions of the accretion flows are traced by different molecules, just as in HL Tau \citep{Yen2017AA...608A.134Y,Yen2019ApJ...880...69Y}, where accretion flows are traced in \COb\ and \HCOp. 
Thus, the gaps may be observable in other molecules or transitions. 

\subsection{Degeneracy of the CMU Model}
The CMU model has four key parameters: stellar mass ($M_*$), centrifugal radius ($r_c$) and and the trajectory inclination ($\theta_0$) and azimuthal angle of apastron ($\phi_0$).
Various combinations of these properties could potentially lead to the same outcome in the fitting percentage calculated in Section \ref{sec:infall}.
We test this by re-running our fitting calculations using the star ($0.24$\msun) + disk ($0.1$\msun) mass and finding new ranges for $\theta_0$ and $\phi_0$.
For the disk mass, we use the value derived by \citet{Yen2017ApJ...834..178Y}. 
For S1, we can still get 100\% fitting by only changing the mass while keeping the same $\theta_0$ and $\phi_0$ range.
For S2, we can change the range of $\theta_0$ from $111^{\circ}-129^{\circ}$ to $112^{\circ}-131^{\circ}$ and $\phi_0$ from $90^{\circ}-117^{\circ}$ to $92^{\circ}-119^{\circ}$.
Doing this we can achieve a fitting of 100\%, like before.
For S3, we can actually increase the fitting percentage from 96.80\% to 100\% by adjusting the range of $\theta_0$ from $125^{\circ}-157^{\circ}$ to $123^{\circ}-155^{\circ}$ and $\phi_0$ from $298^{\circ}-368^{\circ}$ to $316^{\circ}-373^{\circ}$.
For S4, the fitting percentage decreases from 29.94\% to 24.85\%.
This is because adding more mass in the center increases the velocity along the trajectories, thus increasing the line-of-sight velocity in our model, and ends up with fitting less of the lower velocity points. 
Adjusting the range of $\theta_0$ and $\phi_0$ had no affect on this value. 
For S5, we can change the range of $\theta_0$ from $31^{\circ}-74^{\circ}$ to $32^{\circ}-81^{\circ}$ and $\phi_0$ from $201^{\circ}-252^{\circ}$ to $190^{\circ}-252^{\circ}$.
This increases the fitting percentage from 99.06\% to 99.15\%.

\citet{Tachihara2007ApJ...659.1382T} estimate the mass enclosed within 4200\au\ of Lupus 3-MMS to be 0.52\msun. 
Therefore, we also test the extreme case by running the model using the star ($0.24$\msun) + envelope ($0.52$\msun) mass and again finding new ranges for $\theta_0$ and $\phi_0$. 
We find that we can achieve a 100\% fit for all the structures, with the exception of S4, which we could only achieve a 7.59\% fit. Again, the low fitting percentage of S4 is due to the larger mass increasing the velocity along the trajectories, which cannot fit the low velocity part of the structure. 
We find $\theta_0$ and $\phi_0$ ranges to be the same as before for S1, $117^{\circ}-135^{\circ}$ and $98^{\circ}-125^{\circ}$ for S2, $119^{\circ}-150^{\circ}$ and $338^{\circ}-381^{\circ}$ for S3, and $34^{\circ}-58^{\circ}$ and $195^{\circ}-253^{\circ}$ for S5.
Our test show that values of $M_*$ between 0.24\msun\ and 0.76\msun\ still find good solutions with the CMU model.
This highlights that the overall results are not so sensitive to $M_*$ when trying to match the trajectories. 

\begin{deluxetable*}{lccccccc}[htb!]
\tablewidth{\textwidth}
\tablecaption{Comparison with Previously Observed Accretion Flows}
\label{tab:compare}
\tablehead{\vspace{-0.1cm}
Source & Class & \colhead{$r_{\mathrm{flow}}$} & \colhead{$j$} & \colhead{$M_{\mathrm{flow}}$} & \colhead{$t_{\mathrm{acc}}$} & \colhead{$\dot{M}_{\mathrm{disk}}$} & \colhead{Ref.}\\
Name & & \colhead{(au)} & \colhead{(\jsunit)} & \colhead{(M$_{\odot}\,$)} & \colhead{(yr)} & \colhead{(M$_{\odot}\,$yr$^{-1}$)} &
}
\startdata 
Lupus 3-MMS & 0 & $591-1,416$ & $2.8-7.0\times10^{-4}$ & $1.7-5.1\times10^{-3}$ & $1.5-9.8\times10^{3}$ & $0.5-1.1\times10^{-6}$ & 1\\
VLA 1623 & 0 & $1,200-3,600$ & $1.0-1.5\times10^{-3}$ & $2.7-6.4\times10^{-3}$ & $4.4-8.7\times10^{3}$ & $1.2\times10^{-6}$ & 2\\
IRAS 03292+3039  & 0 & $10,500$ & - & 0.1 & $9.6\times10^{4}$ & $1\times 10^{-6}$ & 3\\
L1489-IRS & I & $2,000-5,000$ & $4.8\times10^{-3}$ & $4-7\times10^{-3}$ & $3-5\times10^{4}$ & $4-7\times10^{-7}$ & 4\\
HL Tau & I & $1,000-2,000 $ & $1.9\times10^{-3}$ & $5.7\times10^{-3}$ & $2.6\times10^{3}$ & $2.2\times10^{-6}$ & 5\\
\enddata
\tablecomments{The values listed for Lupus 3-MMS are only from the accretion flows with a fitting percentage $>$95\%. The radius of the accretion flow ($r_\mathrm{flow}$) is the projected 2D radius on the plane-of-sky. Refs:  (1) This work, (2) \citet{CheongThesis2018}, (3) \citet{Pineda2020NatAs.tmp..151P}, (4) \citet{Yen2014ApJ...793....1Y}, (5) \citet{Yen2017AA...608A.134Y}}
\end{deluxetable*}

\subsection{Accretion Flow Properties and Comparison}
\label{sec:PrevObs}
Our analyses indicate the presence of at least four clear accretion flows that match with the CMU model, the most to be found in any source to date.
In Table \ref{tab:properties}, we derived several properties of the accretion flows found in Lupus 3-MMS. 
Here, we discuss these properties and compare them with previously observed accretion flows. 
Previous observations find accretion flows to be either singular (asymmetric) or bilaterally symmetric. 
To date, there are the only four sources (L1489-IRS, HL Tau, VLA 1623 and IRAS 03292+3039) with observed accretion flow structures.
Table \ref{tab:compare} shows a summary of previously observed accretion flows.
In this section, we discuss the various parameters and compare them to other sources.
When comparing these other sources to Lupus 3-MMS, we will only compare to values in the structures that fit $>$95\% with the CMU model. 

\subsubsection{Flow Mass}
The mass of the accretion flow ($M_{\mathrm{flow}}$) quantifies the reservoir of material available to feed the protostellar disk. This value is given as the H$_2$ flow mass, as it is the most abundant molecule. 
It is calculated by using the column density maps and the area of the structure estimated by the pixel area and the number of pixels.
We then assumed an abundance ratio between \COa\ and H$_2$, then multiplied by the H$_2$ mass to approximate the mass of the entire flow structure. 
We estimate the values in Lupus 3-MMS to be $1.3-5.1\times10^{-3}$\msun.
The order of $\times10^{-3}$\msun\ is overall comparable to the derivations in other accretion flows. 
It is most similar to VLA 1623, which shows slightly higher values of $2.7-6.4\times10^{-3}$\msun, also using \COa.
There is a discrepancy between Lupus 3-MMS and IRAS 03292+3039, where \citet{Pineda2020NatAs.tmp..151P} find a huge flow mass of 0.1\msun.
This difference comes from the length of their flow being much longer, meaning the area of the flow is larger. 
In the two Class I sources, L1489-IRS and HL Tau, the masses are on the higher end compared to Lupus 3-MMS due to the similarities in lengths, but overall they are still consistent.

\subsubsection{Infall Timescale}
The infall timescale ($t_{\mathrm{infall}}$) measures how long it will take the material in the flow to reach the disk. 
This value should be consistent with timescales shorter than that of protostellar evolution, which for low-mass protostars, lasts roughly $1.6\times 10^5\,$yr for the Class 0 phase \citep{Evans2009ApJS..181..321E}.
This value is estimated kinematically using the lengths of the flows and their velocities. 
We estimate values on the order of $10^3\,$yr in Lupus 3-MMS. 
Again, this is comparable to VLA 1623 which was calculated to be $4.4-8.7\times10^3\,$yr.  
In contrast, \citet{Pineda2020NatAs.tmp..151P} estimate the infall timescale in IRAS 03292+3039 using a different method based on the free-fall timescale ($t_\mathrm{ff}$), which depends on the radius of the envelope and the mass enclosed within that radius. 
We use the mass enclosed derived by \citet{Tachihara2007ApJ...659.1382T}, a value of 0.52\msun.
Their value is estimated to be much higher due to the higher envelope mass and larger radius. 
We estimate the free-fall time scale in Lupus 3-MMS to compare with IRAS 03292+3039 and the values derived kinematically.
Using the same method as in \citet{Pineda2020NatAs.tmp..151P}, we calculate $t_\mathrm{ff}$ to be $6.01\times10^4\,$yr, comparable to the value of $9.6\times10^4\,$yr for IRAS 03292+3039. 
L1489-IRS is on an order of magnitude of $\sim10^4\,$yr, while HL Tau is $\sim10^3\,$yr. 
The infall timescale is similar in these sources and may be reflected by the initial size-scale of the envelope. 

\subsubsection{Mass-Infall Rate}
The mass-infall rate ($\dot{M}_{\mathrm{infall}}$) assesses the rate at which material falls onto the protostellar disk from the infalling accretion flows. 
This value is estimated by dividing the mass of the flows with the infall timescale. 
For the Class 0 sources these values are comparable, all around $\sim10^{-6}$\mdotunit. 
For the Class I sources, it varies from $10^{-7}$\mdotunit\ to $10^{-6}$\mdotunit\ in L1489-IRS and HL Tau, respectively. 
One reason the mass-infall rate may vary between sources is due to the remaining envelope mass. 
Sources with more envelope mass are expected to be accreting more material, and vice-versa. 
From the Class I sample, envelope masses were derived from IRAM 30m dust continuum observations to be $0.03$\msun\ and $0.13$\msun\ for L1489-IRS and HL Tau at 4200\au, respectively \citep{Motte2001A&A...365..440M}.
This explains the discrepancy between the accretion rates in the two Class I sources, as HL Tau has a much higher value.
In Lupus 3-MMS, we previously mention this value was derived to be 0.52\msun\ at 4200\au. 
The envelope mass is substantially larger due to the younger nature of our source. 
In VLA 1623, \citet{Murillo2018A&A...615L..14M} derive an envelope mass of $\sim1.0$\msun, assuming the two binary sources A \& B share the same envelope. 
\citet{Pineda2020NatAs.tmp..151P} cite a dense core mass of 3.2\msun\ for IRAS 03292+3039, the largest of all the sources. 
The higher mass-infall rates in the Class 0 sources all have much higher envelope masses and the lower envelope mass in L1489-IRS has a lower mass-infall rate.
Assuming the previously derived envelope mass in Lupus 3-MMS and a constant accretion rate, it will take around $10^6\,$yr to deplete the remaining envelope mass. 

The mass-infall rate of $\sim1.53\times10^{-6}$\mdotunit\ for a $10\,$K singular isothermal sphere \citep{Shu1977ApJ...214..488S} is consistent with the values derived for Lupus 3-MMS.
In the context of numerical simulations, most show the presence of extended accretion flow structures \citep[e.g.,][]{Li2014ApJ...793..130L,Seifried2015MNRAS.446.2776S}. 
In more complicated MHD simulations, only a handful derive the mass-infall rate from the envelope to the disk. 
\citet{Machida2016MNRAS.463.4246M} study the formation of the circumstellar disk in strongly magnetized cores. 
They derive mass-infall rates of $\sim10^{-5}$\mdotunit.
\citet{Lee2021A&A...648A.101L} study disk formation in a magnetized core with ambipolar diffusion. 
They find mass-infall rates onto the disk to be $\sim10^{-6}$\mdotunit\ at both 40 and 80$\,$kyr after the formation of the sink particle. 
One difference between these two simulations was the inclusion of ambipolar diffusion in the latter, which helps to reduce the efficiency of magnetic braking in the envelope. 
Our model does not include magnetic fields, but the mass-infall rate is still roughly consistent. 
Magnetic field observations are needed to carry out a more robust comparison with MHD simulations. 

\subsubsection{Specific Angular Momentum}

Angular momentum is important for the formation of the protostellar disk in the earliest stages of star formation. 
The values we calculate for the specific angular momentum ($j$) of the accretion flows in Lupus 3-MMS are on the order of $10^{-4}$\jsunit.
The specific angular momentum at the edge of the rotationally supported disk can be defined as the specific angular momentum of trajectories lying on the disk plane ($j_{\mathrm{d}}=\sqrt{G M_* r_d}=\sqrt{G M_* r_c}\sin\theta_0$ as $\theta_0=90^\circ$; see Equations \ref{eq:5} and \ref{eq:6}), with a value of $7.25\times10^{-4}$\jsunit. 
Compared to other accretion flows, which show $j$ on the order of $10^{-3}$\jsunit, the specific angular momentum in Lupus 3-MMS is roughly an order of magnitude lower.
This is due to the low stellar mass of 0.24\msun, compared to other sources which are more evolved into the Class I stage (L1489-IRS and HL Tau), binary systems where both of the sources are used for the estimation (VLA 1623), and a Class 0 where they use the envelope mass in their model calculation (IRAS 03292+3039).
As the protostellar mass increases over time, then the angular momentum would also increase.
This could suggest Lupus 3-MMS is in the early stages of collapse. 
In comparison to other studies of the specific angular momentum in Class 0 sources, our results are consistent. 
\citet{Pineda2019ApJ...882..103P} find specific angular momentum values to be between $10^{-4}-10^{-3}$\jsunit\ at distances less than 3000\au.
Additionally, \citet{Gaudel2020A&A...637A..92G} find the specific angular momenta of 7 out of 8 Class 0 sources to be $2-7\times10^{-4}$\jsunit.


\subsubsection{Total Angular Momentum Budget}
Estimating the total angular momentum ($J$) that will be delivered to the disk is another way to assess the material being fed to disk scales. 
We take the specific angular momenta of each flow and multiply it by the respective flow mass to calculate a range for each structure. 
If we add up the total angular momentum of each of the accretion flows, we find the total angular momentum in the process of being delivered to the disk to be $1.21-1.61\times10^{-5}$\Jsunit. 
We then calculate the total angular momentum in the disk by taking its specific angular momentum and multiplying it by disk mass (0.1\msun, \citealt{Yen2017ApJ...834..178Y}) to get a value of $7.25\times10^{-5}$\Jsunit. 
This is the same order of magnitude, but slightly less than the disk.
As a check, we also calculate the total angular momentum assuming a disk surface density distribution \citep{Williams2011ARA&A..49...67W}.
We find from this method, the total angular momentum of the disk is $\sim1.00\times10^{-5}$\Jsunit. 
The similarity in total angular momentum between the disk and accretion flows may hint that the disk is relatively young and hasn't had enough time to inherit more angular momentum from the parent envelope. 
Whether the angular momentum of the infalling-rotating envelope can be delivered all the way to the disk depends on the degree of magnetic braking, which can in principle remove angular momentum efficiently \citep[e.g.,][]{Mellon2008ApJ...681.1356M}. 
Dust polarization observations toward this object are needed to understand the magnetic field morphology and strength, which would be useful in understanding the role of magnetic braking in this object. 
Some angular momentum can also be removed by the protostellar outflows, but we do not have clear detection of outflow rotation in our current data. 
If material is falling in from everywhere around the disk, and the flows we detect are just over-densities of that material, then missing flux from shorter baselines could also potentially increase the amount of angular momentum being delivered to the disk.

\subsection{The Origin of Accretion Flows}
Observations of different molecules give us an insight into where the accretion flows originate. 
\citet{Pineda2020NatAs.tmp..151P} find their 10,500\au\ asymmetric accretion flow structure is traced by the ``chemically fresh" carbon-chain molecules HC$_{3}$N, CCS and $^{13}$CS, and not by more chemically evolved molecules like \NNHp\ or \NNDp \citep{Bergin2007ARA&A..45..339B, Sakai2013ChRv..113.8981S}. 
They conclude from this that the material comes from beyond the parental denser core ($>$10,000\au).
As for Lupus 3-MMS, the flows are traced by \COa, a molecule present throughout the star formation process.
\NNDp\ was observed but not detected in these archival observations. 
In order to accurately assess the possible chemical complexity of these flows, observations of other fresh and evolved molecules are needed.
Additionally, throughout this paper we have seen that the addition of shorter baseline observations could help fill in the missing flux to further investigate the size and structure of the accretion flows on larger scales. 
Therefore, the present data is not enough to conclude whether these structures originate from the dense core or from beyond.

\section{Conclusion} \label{sec:conclusion}
Identifying accretion flows is difficult, but important for characterizing how material is transported to the protostellar disk.
We have re-analyzed ALMA archival observations of a young Class 0 protostar, Lupus 3-MMS, in \COan, \COcn, \SOn, and \continuum\ emission.
We uncover the dynamics of several accretion flows using the CMU model and demonstrate a case where accretion flows are hiding in the so-called ``outflow cavity''. 
The main results are summarized as follows:
\begin{enumerate}
  \setlength\itemsep{0.5em}
  \item We identify extended accretion flow-like structures in \COa\ along the sides of the red and blue-shifted \COc\ outflows after re-cleaning the data with a larger beamsize. 
  The extended structures range in lengths of $5^{\prime\prime}-15^{\prime\prime}$, while showing various line-of-sight velocity components and low velocity dispersion. 
  The disk is found in the central region, as indicated by the high-velocity \COa\ red and blue-shifted components corresponding with the continuum emission and high velocity dispersion. 
  The \SO\ emission near the disks indicates the possible presence of accretion shocks. 
  
  \item Comparing the velocity of the extended \COa\ structure with the outflow emission yields no similarities in PV space. 
  Making cuts in the disk plane at different points along the outflows and comparing with a globalized case of our CMU model shows the \COa\ emission to match well with our infalling model curves, while the outflow shifts to higher velocities at greater distances from the source. These different dynamics are important for disentangling emission between the outflows and infalling gas.
  
  \item Based on our dendrogram analysis of five coherent components within the ex-tended structures, we identify four structures to be accretion flows that fit remarkably well with the CMU model with fitting percentages of 95\%. 
  We translate the matching points from PPV space to PPP space and find that the accretion flows are separated from the outflow in 3D space. This highlights that studying emission near the outflow cavities is important for understanding the dynamics in young protostellar systems. 

  \item Using the matching points of these four structures, we derive specific angular momenta of $2.8-7.0\times10^{-4}$\jsunit, column densities of $1.2-3.8\times10^{-12}\,\mathrm{cm^{-2}}$, flow masses of $1.7-5.1\times10^{-3}$\msun, infall timescales of $1.5-9.8\times10^{3}\,\mathrm{yr}$ and mass-infall rates of $0.5-1.1\times10^{-6}$\mdotunit. 
  We find that Lupus 3-MMS is comparable to other Class 0 sources with detected accretion flows. The higher mass-infall rates in the Class 0 sources show the importance of accretion at this phase, when the object is still highly embedded in its parent envelope. The mass-infall rate is comparable to MHD simulations, indicating that such a simplified model can be a good approximation to compare to observations. 
\end{enumerate}

\acknowledgments
T.J.T. and S.-P.L. acknowledge support from the
Ministry of Science and Technology of Taiwan with grant
MOST 106-2119-M-007-021-MY3 and MOST 109-2112-M-007-010-MY3.
C.-F.L. acknowledges grants from the Ministry of Science and Technology of Taiwan (MoST 107-2119-M-001-040-MY3) and the Academia Sinica (Investigator Award AS-IA-108-M01).
H.-W.Y. acknowledges support from MOST 108-2112-M-001-003-MY2 and MOST 110-2628-M-001-003-MY3.
Z.-Y.L. is supported in part by NASA 80NSSC20K0533 and NSF AST-1716259.
This work used high-performance computing facilities operated by the
Center for Informatics and Computation in Astronomy (CICA) at National
Tsing Hua University. 
This equipment was funded by the Ministry of
Education of Taiwan, the Ministry of Science and Technology of Taiwan,
and National Tsing Hua University. 
This paper makes use of the following ALMA data: ADS/JAO.ALMA\#2013.0.00879.S. ALMA is a partnership of ESO (representing its member states), NSF (USA) and NINS (Japan), together with NRC (Canada), MOST and ASIAA (Taiwan), and KASI (Republic of Korea), in cooperation with the Republic of Chile.
The Joint ALMA Observatory is operated by ESO, AUI/NRAO and NAOJ.

\vspace{5mm}
\facilities{ALMA}
\software{Astropy \citep[][\url{http://astropy.org}]{astropy1, astropy2}, Astrodendro \citep[][\url{http://dendrograms.org/}]{Rosolowsky2008ApJ...679.1338R, astrodendro}, CASA \citep[v5.1.2+4.6.2, ] [\url{http://casa.nrao.edu/}]{CASA}, Matplotlib \citep[][\url{http://matplotlib.org/}]{matplotlib}, Numpy \citep[][\url{http://numpy.org/}]{numpy}}

\clearpage
\bibliography{references}
\bibliographystyle{aasjournal}

\appendix
\counterwithin{figure}{section}
\counterwithin{table}{section}

\section{Does the CMU Model Produce Streams?} \label{sec:streams}
\begin{figure}[t]
\centering
\includegraphics[trim=10cm 0cm 14cm 1cm, clip,  width=\textwidth]{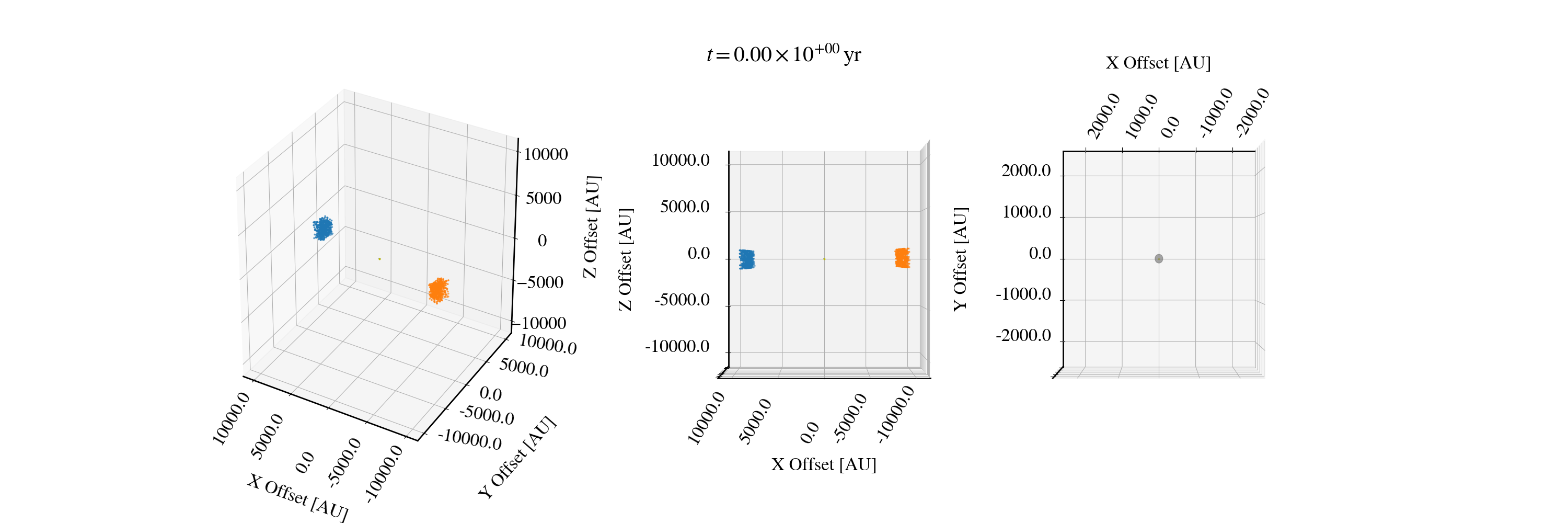}
\includegraphics[trim=10cm 0cm 14cm 1cm, clip, width=\textwidth]{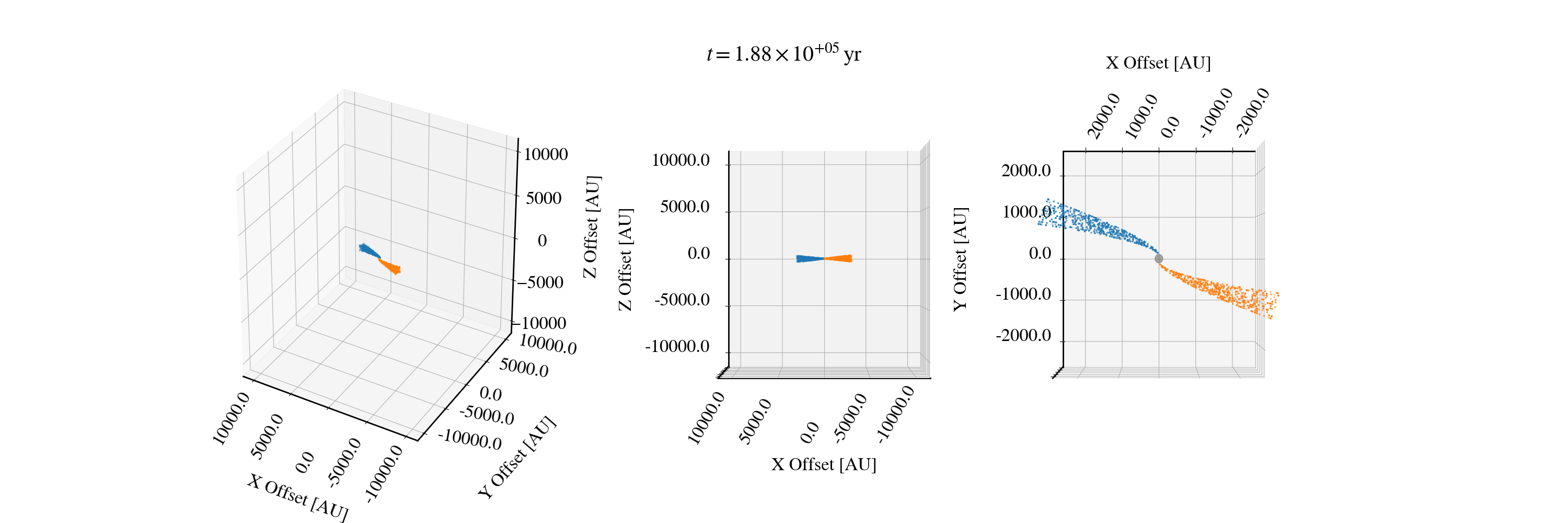}
\caption {CMU Model Simulation 1 of infalling particles around a protostar at $t=0\,$yr (top) and the final time at which all of the particles radii are greater than the centrifugal radius, $r_c$ (bottom). The initial and final times are shown above each figure, respectively. The left panel shows an overall 3D view of the system, the middle panel shows an edge-on view and the right panel shows a zoomed-in face-on view. The $x$ and $z$ axes represent the plane-of-sky axes and the $y$ axis represents the line-of-sight axis. The gray circles in the right-most plots indicate the centrifugal radius. An animation of this figure is available. It starts at t=0$\,$yr and ends at at 1.88e+05$\,$yr. The sequence repeats 3 times and the real-time duration of the animation is 24 seconds.}
\label{fig:cmu_sim_1}
\end{figure}

\begin{figure}[t]
\centering
\includegraphics[trim=10cm 0cm 14cm 1cm, clip,  width=\textwidth]{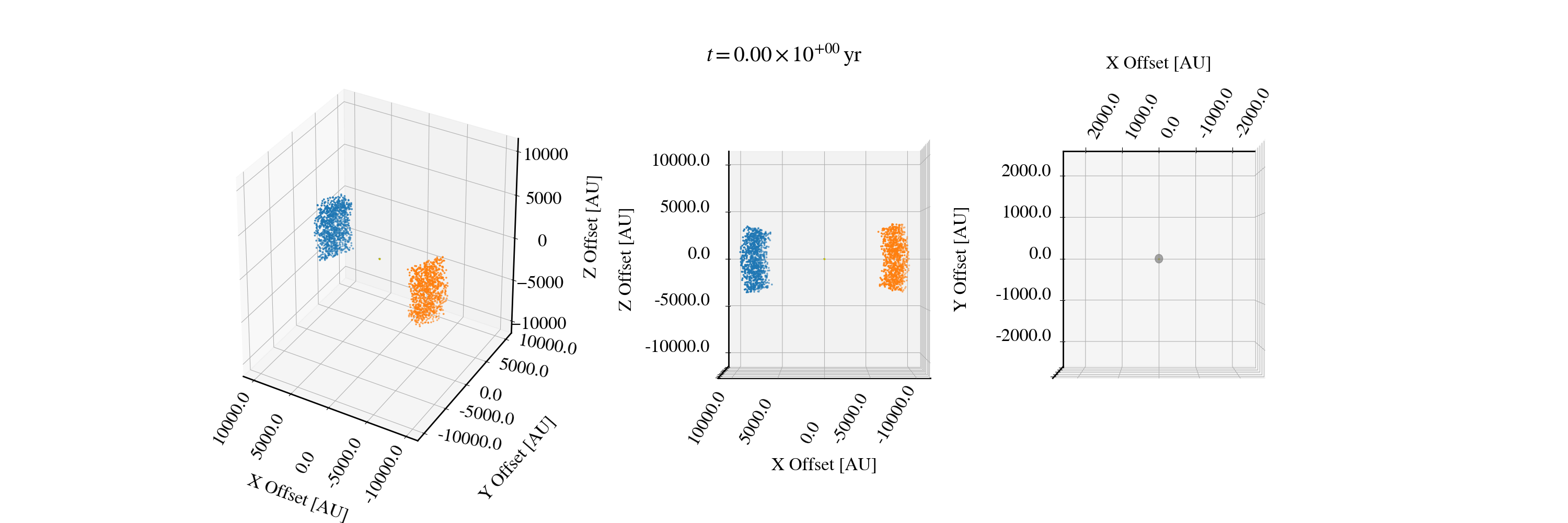}
\includegraphics[trim=10cm 0cm 14cm 1cm, clip, width=\textwidth]{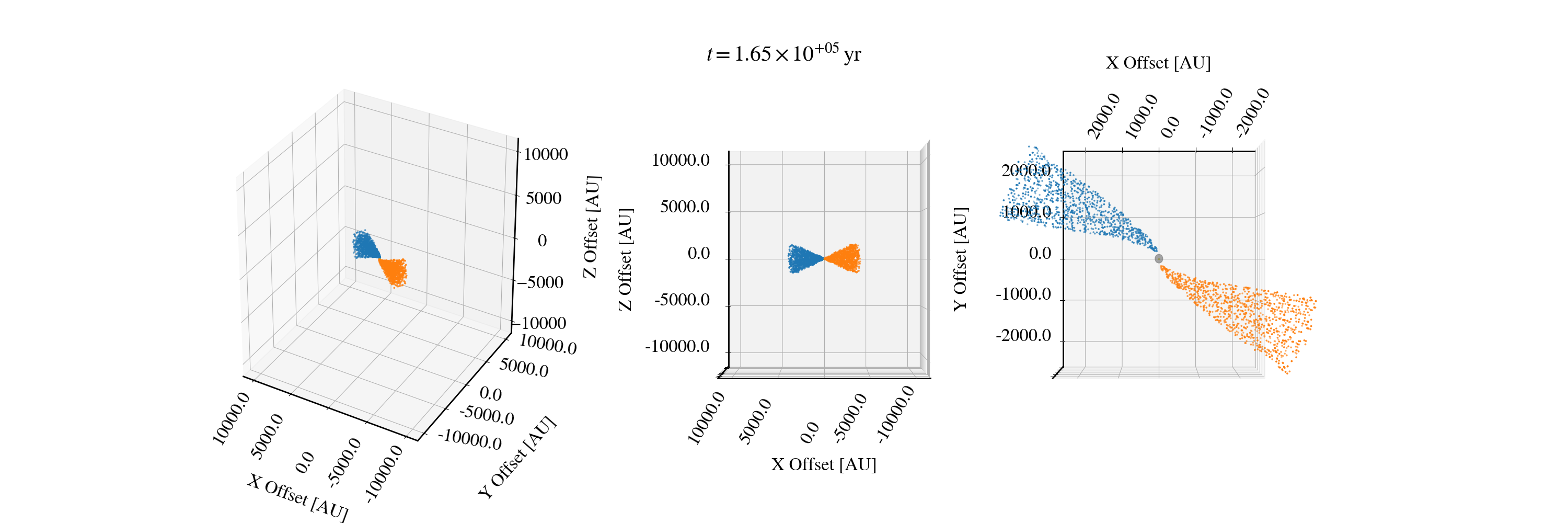}
\caption {Same as Figure \ref{fig:cmu_sim_1} but for Simulation 2. An animation of this figure is available. It starts at t=0$\,$yr and ends at at 1.65e+05$\,$yr. The sequence repeats 3 times and the real-time duration of the animation is 24 seconds.}
\label{fig:cmu_sim_2}
\end{figure}

\begin{figure}[t]
\centering
\includegraphics[trim=10cm 0cm 14cm 1cm, clip,  width=\textwidth]{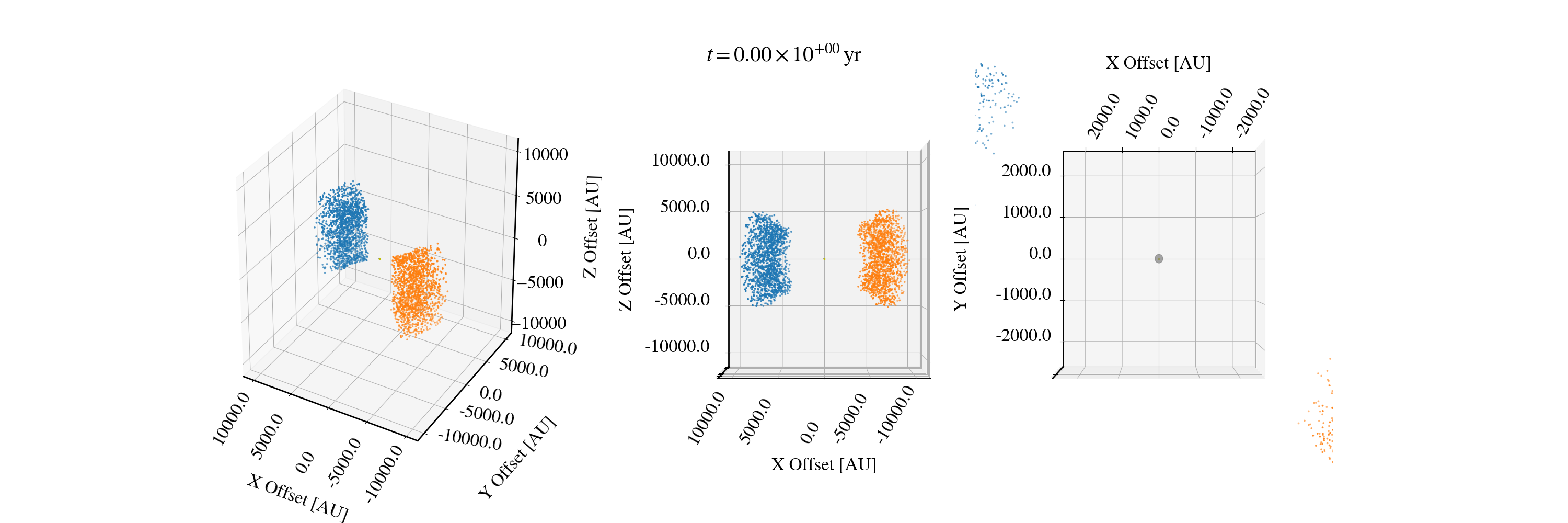}
\includegraphics[trim=10cm 0cm 14cm 1cm, clip, width=\textwidth]{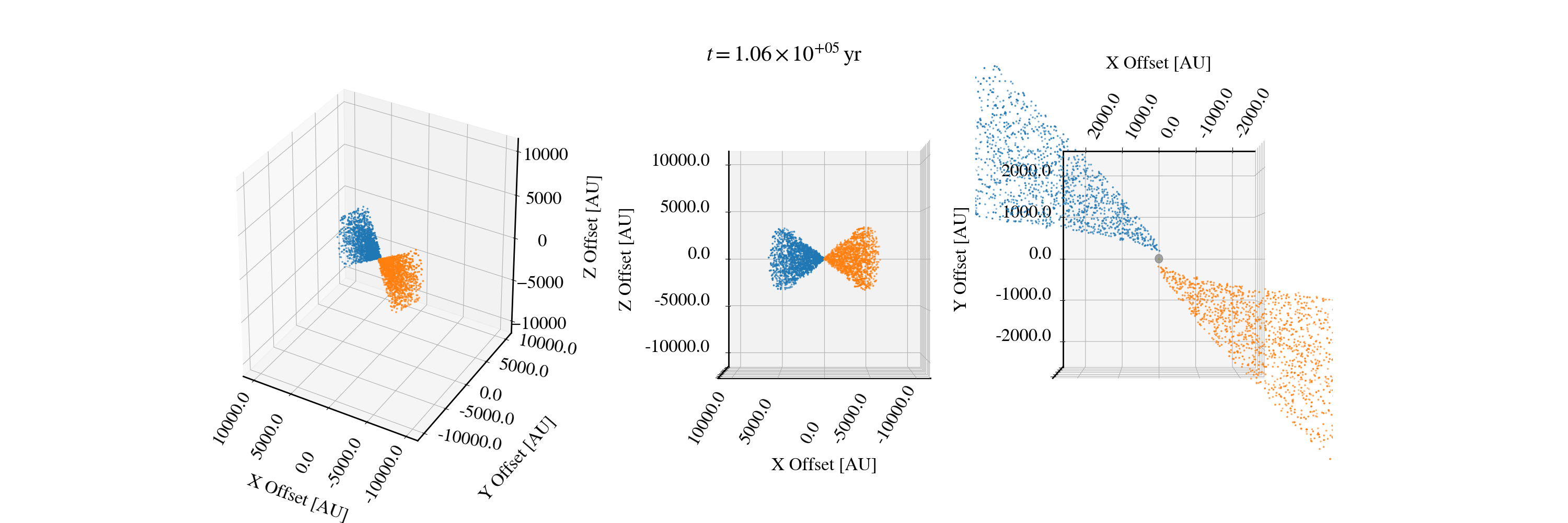}
\caption {Same as Figure \ref{fig:cmu_sim_1} but for Simulation 3. An animation of this figure is available. It starts at t=0$\,$yr and ends at at 1.06e+05$\,$yr. The sequence repeats 3 times and the real-time duration of the animation is 24 seconds.}
\label{fig:cmu_sim_3}
\end{figure}

The CMU model produces the trajectory path of a particle over a certain amount of time, while in reality, observational data gives us a snapshot in time. 
Therefore, one important question we need to answer is whether or not the CMU model can produce a stream like structure from a patch of infalling particles. 
To do this, we created three `simulations' of various sized clumps of randomly distributed particles by changing the values of $r_0$, $\theta_0$ and $\phi_0$ (Table \ref{tab:simulations}). 
The stellar (point) mass ($M_*$) and centrifugal radius ($r_c$) are kept at a constant for each simulation. 
We increase the number of particles ($N_\mathrm{particles}$) as we increased the size of the initial clumps. 
We ran each simulation until the final time at which all of the particles were greater than the centrifugal radius, $r_c$. 
All of our simulations produce clear streams when the clumps of particles are about to collide with the disk (see Figures \ref{fig:cmu_sim_1}, \ref{fig:cmu_sim_2} and \ref{fig:cmu_sim_3}). 
This is due to the fact that since the gravitational force at inner radii is greater than at the outer radii, the velocity at the inner radii will increase faster and stretch out the clump as it gets closer to the disk to produce a stream. 
The streams vary in size based on the initial size of the clump. 

\begin{deluxetable}{ccccccccc}[htb!]
\tablewidth{\textwidth}
\tablecaption{CMU Simulations for Different Sized Clumps of Particles}
\label{tab:simulations}
\tablehead{\vspace{-0.1cm}
& \colhead{$r_0$ range} & \colhead{$\theta_0$ range} & \colhead{$\phi_0$ range} & \colhead{$M_{*}$}& \colhead{$r_c$} & \colhead{$N_{\mathrm{particles}}$} & \colhead{$t_{\mathrm{final}}$} & \colhead{Streams?}\\
& \colhead{(au)} & \colhead{($^{\circ}$)} & \colhead{($^{\circ}$)} & \colhead{($M_{\odot}$)}& \colhead{(au)} & & \colhead{($\times10^{5}\,$yr)} &
 }
\startdata 
Simulation 1 & 8500--9800 & 85--95 & 0--10, 180--190 & 0.1 & 100 & 500 & 1.88 & Yes\\
Simulation 2 & 7800--9800 & 70--110 & 0--25, 180--205 & 0.1 & 100 & 1000 & 1.65 & Yes\\
Simulation 3 & 5800--9800 & 60--120 & 0--35, 180--215 & 0.1 & 100 & 1500 & 1.06 & Yes\\
\enddata
\tablecomments{The columns of $r_0$, $\theta_0$ and $\phi_0$ represent the initial radius of the particle, polar angle range of the particles and azimuthal angle range of the particles. For $\phi_0$, the first range corresponds to the orange particles, while the second range corresponds to the blue patch of particles.$M_*$ and $r_c$ represent the point mass and centrifugal radius, respectively. $N
_\mathrm{particles}$ is the number of particles used in the simulation. $t_\mathrm{final}$ is he final time at which all of the particles radii are greater than the centrifugal radius,$r_c$. For each simulation, we label whether or not streams are clearly seen.} 
\end{deluxetable}

\section{PV Diagrams} \label{sec:pvcomp}

\begin{figure*}[ht!]
\centering
\includegraphics[trim=0cm 0cm 0cm 0cm, clip, width=\textwidth]{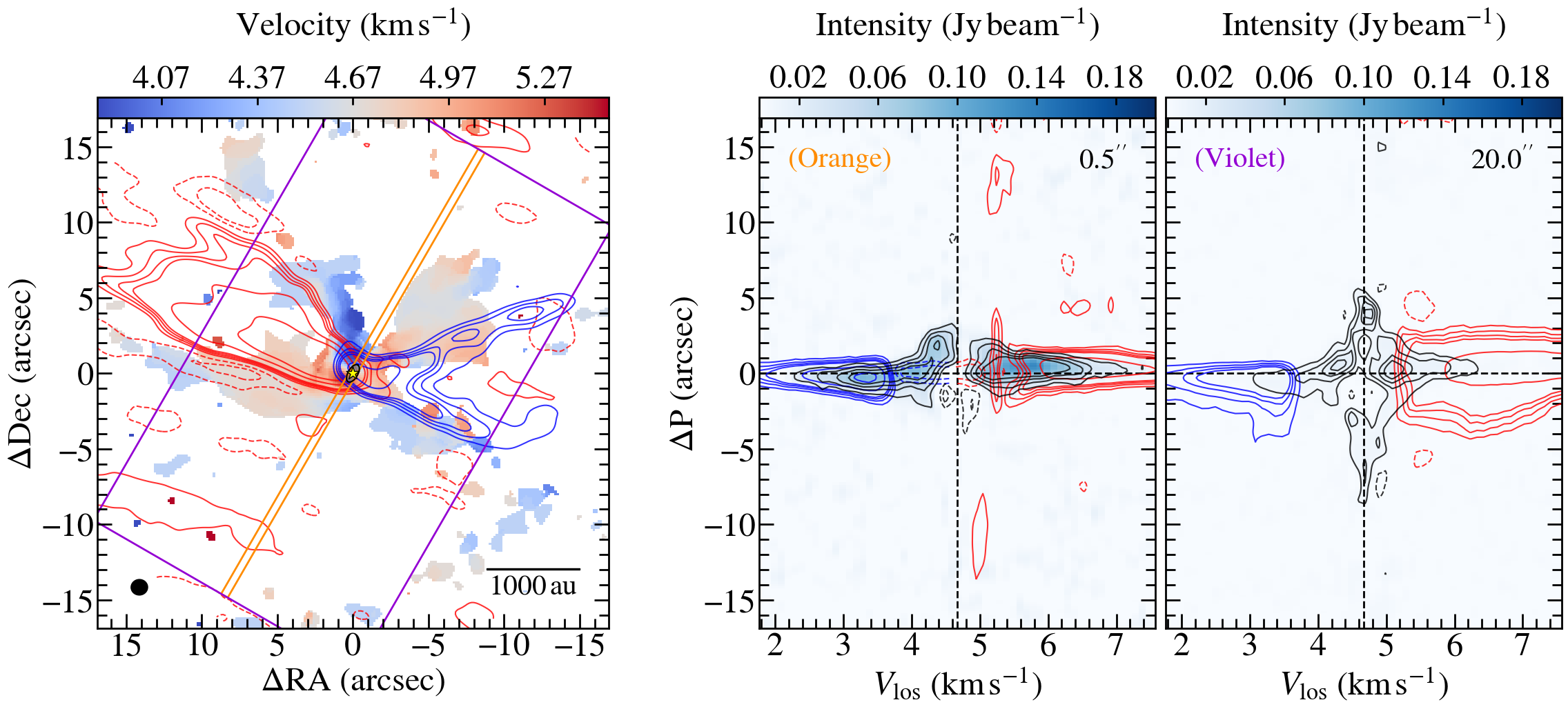}
\caption {Position-Velocity (PV) cuts of our \COa\ data along the disk axis. The PA of disk axis is $150^{\circ}$. We show the positions and sizes of the PV cuts overlaid on the intensity-weighted velocity map (left). The two widths of the cuts are 0.5\arcsec\ (orange) and 20.0\arcsec\ (purple), both with lengths of 35.0\arcsec. We show the \COa\ PV diagrams (center and right) for each cut width with black contours representing levels of -10, -5, 5, 10, 15, 20, 50 and 100$\sigma$, where $\sigma=3.59$ (0.5\arcsec) and $0.91$ (20.0\arcsec) $\mathrm{mJy\,beam^{-1}}$. We over-plot the \COc\ contours in red and blue for the red and blue-shifted outflows, respectively, with the same contour levels as in \COa, but $\sigma=11.04$ (0.5\arcsec) and $6.20$ (20.0\arcsec) $\mathrm{mJy\,beam^{-1}}$. All $\sigma$ values were calculated from an square region in the PV image with no emission.}
\label{fig:c18o_pvdisk}
\end{figure*}

\begin{figure*}[ht!]
\centering
\includegraphics[trim=0cm 0cm 0cm 0cm, clip, width=\textwidth]{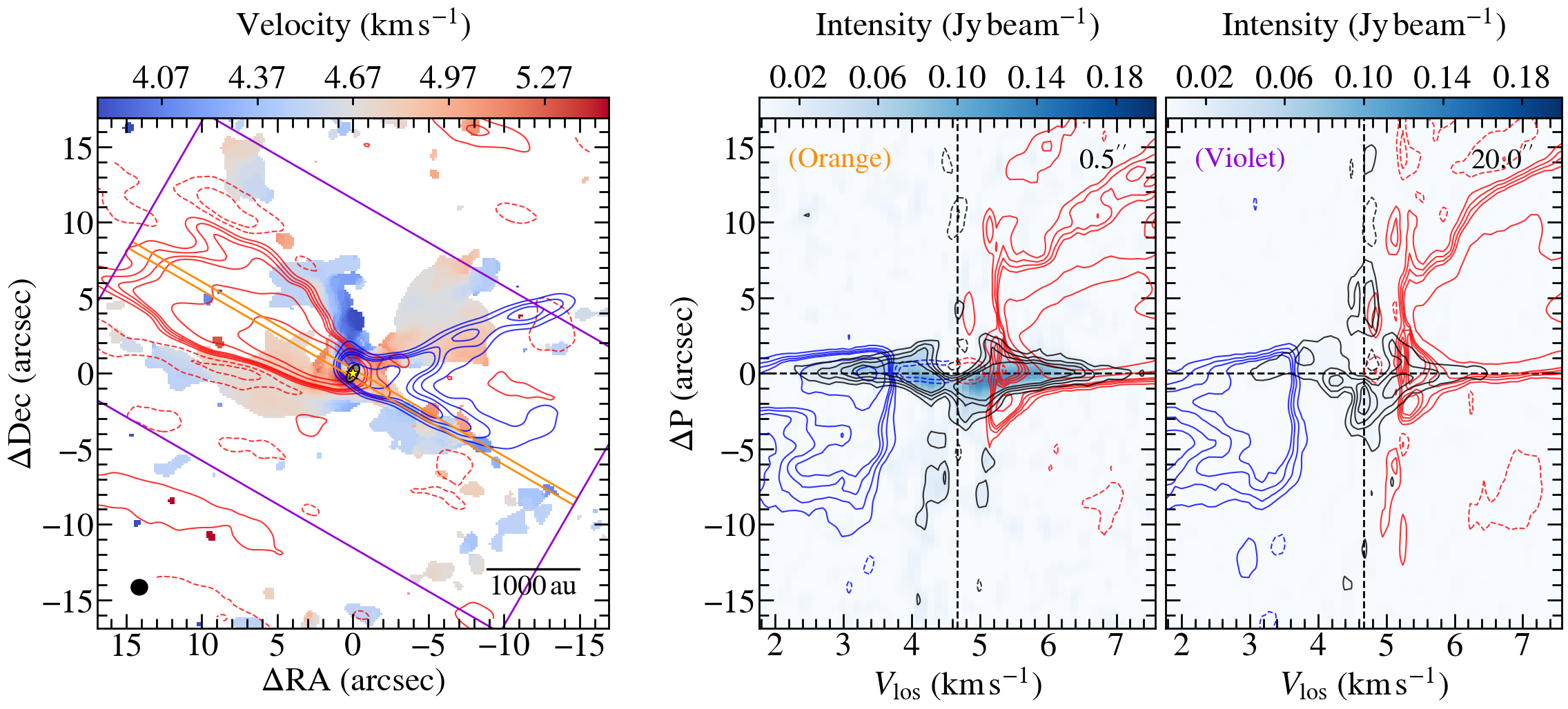}
\caption {Same as Figure \ref{fig:c18o_pvdisk}, but the cut is taken along the outflow axis. Although the PA of both outflows is given by \citet{Yen2017ApJ...834..178Y}, we just take the PA to be the PA of the disk axis minus $90^{\circ}$ for simplicity. For the \COa\ contours, $\sigma=3.19$ (0.5\arcsec) and $0.84$ (20.0\arcsec) $\mathrm{mJy\,beam^{-1}}$. For the \COc\ contours, $\sigma=9.48$ (0.5\arcsec) and $2.33$ (20.0\arcsec) $\mathrm{mJy\,beam^{-1}}$.}
\label{fig:c18o_pvoutflow}
\end{figure*}

To initially assess whether the \COa\ emission is connected to the outflow, we make PV cuts parallel and perpendicular to the disk.
We produce cuts of 0.5\arcsec\ and 20.0\arcsec\ for each case (Figures \ref{fig:c18o_pvdisk} and \ref{fig:c18o_pvoutflow}). 
For our cuts along the disk axis (PA=$150^{\circ}$), the velocity structure between the two molecules becomes more noticeable in the larger cut.
In the 0.5\arcsec\ cut, the positional extent is similar between \COa\ and \COc, with the velocity structure overlapping between the two molecules. 
When we increase the cut to 20.0\arcsec, we find the \COa\ emission to resemble a diamond like shape, indicative of infalling and rotational motion \citep[see][]{Sakai2014Natur.507...78S}.
On the contrary, the \COc\ emission in this cut resembles an elongated structure which is more extended along the velocity axis.
For our cuts along the the outflow axis (PA=$60^{\circ}$), the velocity structure in both cuts shows differences between the molecules. 
In the 0.5\arcsec\ cut, the \COa\ emission is found in the central part of the PV diagram, while the \COc\ emission increases in velocity as you move further away from the center. 
When we increase the cut to 20.0\arcsec, we see the same idea that most of the \COa\ emission is centralized, while the \COc\ emission increases in velocity as you move further away. 
In this figure though, it seems that some of the \COa\ emission looks to follow the edges of the outflows on the red-shifted side. 
Therefore, we think that this method of disentangling the \COa\ and \COc\ emission can be more confused and less robust. 

\clearpage
\section{Column Density Maps} \label{sec:cdm}
\begin{figure*}[ht!]
\centering
\includegraphics[trim=0cm 0cm 0cm 0cm, clip, width=0.97\textwidth]{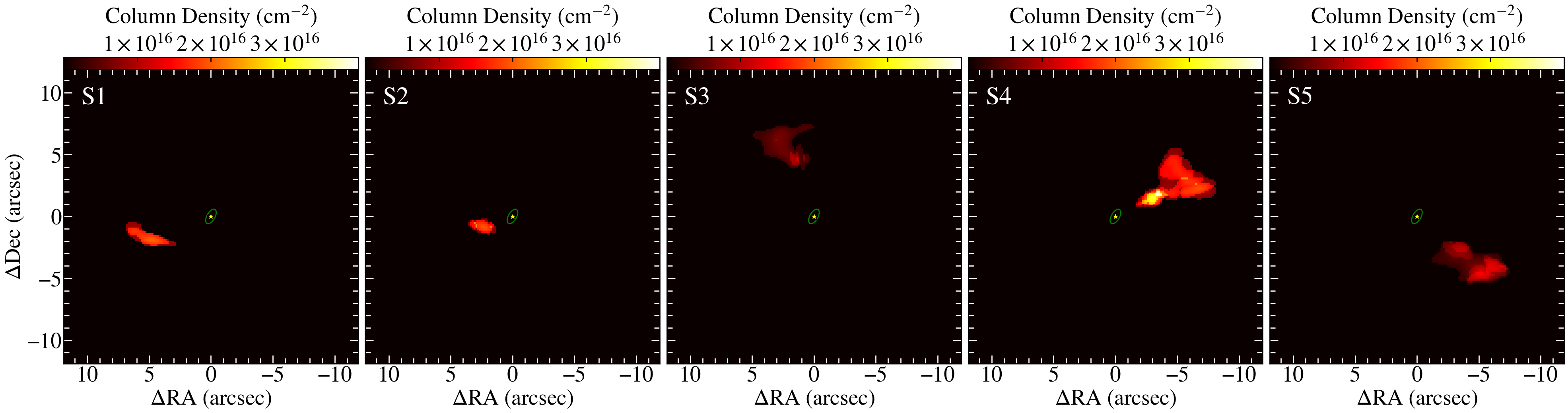}
\caption {Column Density maps for each dendrogram structure. The central source is marked by the yellow star and the rotationally-supported disk is marked by the green circle.}
\label{fig:c18o_cdm}
\end{figure*}

We produce \COa\ column density maps under the local thermal equilibrium (LTE) assumption using the equations and values given in \citet{Mangum_2015}.
We first make integrated-intensity maps of our five dendrogram structures. 
Then, we use the values from those maps and calculate the LTE column density via

\begin{equation*}
    N_{\mathrm{tot}}=\frac{3c^2}{16 \pi^3 \Omega_s S \mu^2 \nu^3}\left(\frac{Q_{\mathrm{rot}}}{g_J g_K g_I}\right)\exp\left(\frac{E_u}{kT_{ex}}\right)\int S_\nu \Delta v,
\end{equation*}

where where $\Omega_s$ is the beam solid angle assuming a Gaussian-shaped source in sr ($\Omega_s=1.133\theta_{\mathrm{maj}}\theta_{\mathrm{min}}$, where $\theta_{\mathrm{maj}}$ \& $\theta_{\mathrm{in}}$ are in rad), $S$ is the line strength (for a linear molecular transition from $J_u\rightarrow J_u-1$, $S=\frac{J_u^2}{J_u(2J_u+1)}$, where $J_u$ is the upper rotational quantum number), $\mu$ is the molecular dipole moment in Debye (1 Debye = $10^{-18}$ statC$\cdot$cm, and 1 statC = 1 cm$^{3/2}\cdot$ g$^{1/2}\cdot$ s$^{-1}$), $\nu$ is the rest frequency of the molecule in GHz, $Q_{\mathrm{rot}}$ is the rotational partition function (for diatomic linear molecules, can be approximated as $Q_{\mathrm{rot}}\simeq\frac{kT}{hB_0}+\frac{1}{3}$), $B_0$ is the rigid rotor rotation constant in MHz, $g_J$ is the rotational degeneracy due to the projection of the angular momentum on the spatial axis $z$ ($g_J=2J_u+1$), $g_K$ is the $K$ degeneracy associated with the internal quantum number $K$ in symmetric and asymmetric top molecules due to projections of the total angular momentum onto a molecular axis ($g_K=1$ for all linear and asymmetric top molecules), $g_I$ is the nuclear spin degeneracy takes account of the statistical weights associated with identical nuclei in a nonlinear molecule with symmetry ($g_I=\frac{g_\mathrm{nuclear}}{g_n}$ and $g_I=1$ for linear molecules), $E_u$ is the upper energy level in Kelvin, $T_{ex}$ is the excitation temperature in Kelvin, $S_\nu$ is the flux density in Jy and $\Delta v$ is the velocity resolution in km/s.
We assume $T_{ex}$ to be $30\,$K, the temperature of the dust continuum emission recently derived by \citet{Vazzano2021A&A...648A..41V}.
For \COa, we use 

\begin{equation*}
    B_{0} = 57635.96\,\mathrm{MHz}
\end{equation*}
\begin{equation*}
    T_{ex} = 30\,\mathrm{K}
\end{equation*}
\begin{equation*}
    \mu = 0.11079\,\mathrm{Debye}
\end{equation*}
\begin{equation*}
    J_u = 2
\end{equation*}
\begin{equation*}
    g_J = 2J_u+1
\end{equation*}
\begin{equation*}
    g_K = 1
\end{equation*}
\begin{equation*}
    g_I = 1
\end{equation*}
\begin{equation*}
    E_u = 15.08\,\mathrm{K}
\end{equation*}
\begin{equation*}
    S = \frac{J_u}{2J_u+1}
\end{equation*}
\begin{equation*}
    \nu = 219.56036\,\mathrm{GHz}
\end{equation*}

to plug into the column density equation and make the maps shown in Figure \ref{fig:c18o_cdm}.

\end{document}